\RequirePackage{fix-cm}

\documentclass[11]{article}
\usepackage{graphicx,authblk}
\usepackage{microtype}
\usepackage{xspace}
\usepackage[colorinlistoftodos]{todonotes}
\usepackage[colorinlistoftodos]{todonotes}
\usepackage{amsmath,amsthm,amssymb}
\usepackage[end]{algpseudocode}
\usepackage{booktabs}
\usepackage{footnote}
\usepackage{url}
\usepackage{tabularx}

\usepackage[todonotes={textsize=footnotesize},final]{changes}
\definechangesauthor[color=blue]{GM}

\usepackage{tikz}
\usetikzlibrary{matrix,positioning,arrows.meta,arrows}
\usetikzlibrary{calc}
\usetikzlibrary{shapes.geometric,arrows}
\usetikzlibrary{decorations.text}
\usetikzlibrary{shapes}
\usetikzlibrary{automata,positioning}

\def\Oh{\mathcal{O}}
\def\eqref#1{(\ref{#1})}
\newcommand{\xx}{\$}
\newcommand{\A}{\Sigma}
\newcommand{\Asize}{\sigma}

\newcommand{\rank}{\mathsf{rank}}
\newcommand{\sel}{\mathsf{select}}

\newcommand{\mergex}{\mathsf{merge}}
\newcommand{\newP}{\ensuremath{\mathsf{newP}}\xspace}

\newcommand{\LCP}{\mathsf{LCP}}
\newcommand{\LCS}{\mathsf{LCS}}

\newcommand{\Count}{C}
\newcommand{\Countx}{C'}

\newcommand{\txx}[1]{\mathsf{s}_{#1}}

\newcommand{\nz}{{n_0}}
\newcommand{\none}{{n_1}}

\newcommand{\oneb}{{\bf 1}}
\newcommand{\zerob}{{\bf 0}}
\newcommand{\Bid}{\mathsf{Block\_id}}
\newcommand{\bid}{\mathsf{id}}

\newcommand{\bv}[1]{Z^{(#1)}}
\newcommand{\kh}{{b(h)}}
\newcommand{\sbot}{0}

\def\Bx{B_{x}}

\algnewcommand\KwTo{\textbf{to }} 
\algnewcommand\KwAnd{\textbf{and }}
\algnewcommand\KwWrite{\textbf{write }}
\algnewcommand\KwTofile{\textbf{to file }}

\renewcommand{\u}{\underline}

\newcommand{\etal}{{\it et al.}\xspace}
\newcommand{\eg}{{\it e.g.}\xspace}

\renewcommand{\S}{\mathcal{C}}

\newcommand{\dbG}{de Bruijn\xspace}
\newcommand{\dbg}{de Bruijn\xspace}
\newcommand{\repr}[1]{\overrightarrow{#1}}
\newcommand{\reprr}[1]{\overleftarrow{#1}}
\newcommand{\last}{\mathsf{last}}
\newcommand{\Wa}{{W}}
\newcommand{\Wx}{{W}^{-}}

\newcommand{\LF}{LF}
\newcommand{\sta}{\mathsf{start}}

\newcommand{\Colz}{\mathcal{C}_0}
\newcommand{\Colo}{\mathcal{C}_1}
\newcommand{\Colzo}{\mathcal{C}_{01}}
\newcommand{\lastz}{\mathsf{last}_0}
\newcommand{\Waz}{{W_0}}
\newcommand{\Gz}{{G_0}}
\newcommand{\Wxz}{W_0^{-}}
\newcommand{\Wxo}{W_1^{-}}
\newcommand{\lasto}{\mathsf{last}_1}
\newcommand{\Wao}{{W_1}}
\newcommand{\Go}{{G_1}}
\newcommand{\Gb}{{G_b}}
\newcommand{\lastx}[1]{\mathsf{last}_{#1}}
\newcommand{\Wax}[1]{{W_{#1}}}
\newcommand{\Wxx}[1]{W_{#1}^{-}}
\newcommand{\laz}[1]{\ell_0(#1)}
\newcommand{\lao}[1]{\ell_1(#1)}

\newcommand{\Bxx}{B_2}
\newcommand{\zzx}{\mathit{0}}
\newcommand{\oddx}{\mathit{1}}
\newcommand{\evenx}{\mathit{2}}
\newcommand{\oox}{\mathit{3}}

\newcommand{\edges}{m}

\def\mysubsubsection#1{\medbreak\noindent\textbf{#1.}}

\def\b#1{\mbox{\boldmath $#1$}}

\newcommand{\langz}{\ensuremath{\mathcal{L}_0}\xspace}
\newcommand{\lango}{\ensuremath{\mathcal{L}_1}\xspace}
\newcommand{\langx}{\ensuremath{\mathcal{L}}\xspace}

\newcommand{\ua}{\ensuremath{\mathcal{U}}\xspace}
\newcommand{\uax}{\ensuremath{\mathcal{U}'}\xspace}
\def\Bx{P}
\def\Bz{S_0} 
\def\Bo{S_1}
\def\Bp{B}
\def\Zp{Z}
\def\Zpx{\Zp'}
\def\Bpx{\Bp'}
\def\Auz{\mathcal{A}_0}
\def\Auo{\mathcal{A}_1}
\def\Aux{\mathcal{A}}
\newcommand{\Wcons}{W-consistent\xspace}
\def\WG{Wheeler graph\xspace}
\def\WGs{Wheeler graphs\xspace}
\def\WA{Wheeler automaton\xspace}
\def\WAs{Wheeler automata\xspace}
\def\WO{Wheeler C-order\xspace}

\newcommand{\lij}[2]{X_{{#1},{#2}}}
\newcommand{\gij}[2]{\neg X_{{#1},{#2}}}
\newcommand{\lab}[1]{\lambda(#1)}
\newcommand{\minmax}{minmax\xspace}

\newcommand{\myset}[1]{\{#1\}}

\newcommand{\inx}{{I}}
\newcommand{\outx}{{O}}
\newcommand{\onex}{\ensuremath{\mathbf{1}}}
\newcommand{\zerox}{\ensuremath{\mathbf{0}}}

\theoremstyle{plain}
\newtheorem{theorem}{Theorem}
\newtheorem{lemma}[theorem]{Lemma}

\theoremstyle{definition}
\newtheorem{definition}[theorem]{Definition}
\newtheorem{property}[theorem]{Property}

\newtheorem{problem}[theorem]{Problem}
\theoremstyle{remark}

\begin{document}

\title{\bf Space efficient merging of \dbG graphs and Wheeler graphs\thanks{Postprint version. The final authenticated version will appear in Algorithmica. A preliminary version of the results in Section~3 appeared in the Proc. 26th Symposium on String Processing and Information Retrieval (SPIRE 2019)~\cite{EgidiLM19}.}
}

\author[1]{Lavinia Egidi}
\author[2]{Felipe A. Louza}
\author[3]{Giovanni Manzini}


\affil[1]{University of Eastern Piedmont, Alessandria, Italy}
\affil[2]{Federal University of Uberl\^andia, Uberl\^andia, Brazil}
\affil[3]{University of Pisa, Pisa, Italy}

\date{}

\maketitle

\begin{abstract}

The merging of succinct data structures is a well established technique for the space efficient construction of large succinct indexes. In the first part of the paper we propose a new algorithm for merging succinct representations of {\em \dbG  graphs}. Our algorithm has the same asymptotic cost of the state of the art algorithm for the same problem {but it uses less than half of its working space.} A novel {important} feature of our algorithm, not found in {any of} the existing tools, is that it can compute the {\em Variable Order} succinct representation of the union graph within the same asymptotic time/space bounds. In the second part of the paper we consider the more general problem of merging succinct representations of {\em \WGs}, a recently introduced graph family which includes as special cases \dbG graphs and many other known succinct {indexes} based on the BWT or one of its variants. \added{In this paper we provide a space efficient algorithm for Wheeler graph merging; our algorithm works under the assumption that the union of the input Wheeler graphs has an ordering that satisfies the Wheeler conditions and which is compatible with the ordering of the original graphs.}

\end{abstract}

\section{Introduction}

A fundamental parameter of any construction algorithm for succinct data structures is its {\em space usage}: this parameter determines the size of the largest dataset that can be handled by a machine with a given amount of memory. Recent works~\cite{latin10j,mics/KarkkainenK17,MuggliBC19,dcc/Siren16} have shown that the technique of building large indexing data structures by merging or updating smaller ones is one of the most effective for designing space efficient algorithms.

In the first part of the paper we consider the {\em\dbG\ } graph for a collection of strings, which is a key data structure for genome assembly~\cite{Pevzner2001}. After the seminal work of Bowe \etal~\cite{wabi/BoweOSS12}, many succinct representations of this data structure have been proposed in the literature (\eg~\cite{almob/ChikhiR13,wabi/AlmodaresiPP17,Belazzougui_2018,dcc/BoucherBGPS15,bioinformatics/MuggliBNMBRGPB17,cpm/BelazzouguiC19}) offering more and more functionalities still using a fraction of the space required to store the input collection uncompressed. 
In this paper we consider the problem of merging two existing succinct representations of {\dbG\ graphs built for} different collections. Since the \dbG\ graph is a lossy representation and from it we cannot recover the original input collection, the alternative to  merging is storing a copy of each collection to be used for building new \dbG\ graphs from scratch. 

Recently, Muggli {\em et al.}~\cite{MuggliBC19,Muggli229641} have proposed a merging algorithm for \dbG\ graphs and have shown the effectiveness of the merging approach for the construction of \dbG\ graphs for very large datasets. The algorithm in~\cite{MuggliBC19} is based on an MSD Radix Sort procedure of the graph edges and its running time is $\Oh(m k)$, where $m$ is the total number of edges and $k$ is the order of the \dbG\ graph. For a graph with $m$ edges and $n$ nodes the merging algorithm by Muggli \etal uses $2(m\log\sigma+m+n)$ bits plus $\Oh(\sigma)$ words of working space, where $\sigma$ is the alphabet size (the working space is defined as the space used by the algorithm in addition to the space used for the input and the output). This value represents a three fold improvement over previous results, but it is still larger than the size of the resulting {succinct representation of the} \dbG graph, which is upper bounded by $2(m\log\sigma+m) + o(m)$ bits.

We present a new merging algorithm that still  runs in $\Oh(m k)$ time, but only uses $4n$ bits  plus $\Oh(\sigma)$ words of working space. For genome collections ($\sigma=5$) our algorithm uses less than half the space of Muggli \etal's: our advantage grows with the size of the alphabet and with the average out-degree $m/n$. Notice that the working space of our algorithm is always less than the space of the resulting {succinct} \dbG graph. Our new merging algorithm is based on a mixed {LSD/MSD  Radix Sort} algorithm which is inspired by the lightweight BWT merging introduced by Holt and McMillan~\cite{bcb/HoltM14,bioinformatics/HoltM14} and later improved in~\cite{spire/EgidiM17,tcs/EgidiM20}. 
In addition to its small working space, our algorithm has the remarkable feature that it can compute as a by-product, with no additional cost, the $\LCS$ (Longest Common Suffix)  between the node labels {in Bowe \etal's representation}, thus making it possible to construct succinct Variable Order \dbG\ graph~\cite{dcc/BoucherBGPS15}, a feature not shared by any other merging algorithm.

In the second part of this paper, we  address the issue of generalizing the results on \dbG graphs, and some previous results on succinct data structure merging~\cite{tcs/EgidiM20,bioinformatics/HoltM14}, to {\it \WGs}. The notion of \WG has been recently introduced in~\cite{tcs/GagieMS17} to provide a unifying view of a large family of compressed data structures loosely based on the BWT~\cite{BW94} or one of its variants. Among the others, the FM-index~\cite{FM05}, the XBWT~\cite{jacm09}, and the BOSS representation of \dbG graphs can all be seen as special cases of (succinct) \WGs. After their introduction, \WGs have become 
objects of independent study and several authors have shown they have some remarkable properties (\eg~\cite{Alanko2020,dcc/AlankoGNB19,dcc/BaierD19,CoPre21,wabi/GGM21,Gibney2019}).

A space efficient algorithm for merging \WGs would automatically provide a merging algorithm for the many practical succinct data structures, present and future, which have the \WG structure. Unfortunately, because of their generality, the problem of merging \WGs \replaced{appears to be}{is} much harder than the problem of merging specific data structures. 
As we discuss in Section~\ref{sec:2sat}, the correct setting for \WG merging is to consider the language $\langx$, defined as the union of the languages recognized by the input graphs when considered as Nondeterministic Finite Automata, and then to build a \WG recognizing $\langx$ (assuming one exists, see~\cite[Lemma 3.3]{AlankoCorr2020}). In this paper we address a slightly simpler problem: we consider the union graph, a graph guaranteed to recognize $\langx$, and ask whether there is an ordering of its nodes that makes it a \WG, \added{with the additional restriction that such ordering must be compatible with the Wheeler orderings of the graphs that contribute to the union. By compatible, we mean that the relative order of the nodes from each single graph are preserved.} Although determining if a graph is a Wheeler graph is NP-complete in the general case~\cite{Gibney2019}, for the special case of the union graph\added{, and with the additional restriction mentioned above,} we show that the problem can be solved in quadratic time via a reduction to the 2-SAT problem. We also describe a space efficient algorithm \added{that is guaranteed to return a Wheeler graph recognizing the language~$\langx$ under the condition that the union graph has a compatible Wheeler ordering (and sometimes even when this condition is not satisfied, as discussed in Section~\ref{ss:iteref}).} If the union graph has $n$ nodes, our algorithm takes $\Oh(n^2)$ time and only uses $4n+o(n)$ bits of working space. A fine point is that sometimes our algorithm does not return the union graph itself but a smaller \WG recognizing $\langx$: this is  not relevant \replaced{for succinct data structures}{in practice} but it is a further indication that the problem should be studied looking at the properties of the union language~$\langx$.

To our knowledge we are the first to tackle the problem of \WGs merging. Although we do not solve this problem in its full generality, our results show that it is possible to perform non trivial operations on a succinct representation of \WGs using a small working space: extending this result would make them even more appealing as general tools for establishing properties of an important class of succinct data structures.

\section{Background and notation} \label{s:notation}

Let $\A = \{ 1,2,\ldots,\Asize\}$ denote the canonical alphabet of size $\Asize$. 
\added{
Let $s[1, n]$ denote a string of length $n$ over $\A$. Given two strings $s_1$ and $s_2$, we write $s_1 \prec s_2$ to denote that $s_1$ is lexicographically smaller than $s_2$.}
Given a string $s[1,n]$ and $c\in\A$, we write $\rank_c(s,i)$ to denote the number of occurrences of $c$ in $s[1,i]$, and $\sel_c(s,j)$ to denote the position of the $j$-th $c$ in~$s$. 
\added{We define $\sel_c(s,0)=0$.}
In this paper we assume a RAM model with word size $w$ with $\Asize=w^{\Oh(1)}$. This ensures that we can represent any string $s$ in $|s|\log\sigma +o(|s|)$ bits, or even $H_0(s) + o(|s|)$ bits, and support $\rank$ and $\sel$ queries in constant time~\cite[Theorem~7]{BNtalg14},
\added{where $H_0(s)$ is the empirical \textit{zero-order} entropy of $s$.}

\subsection{\dbG graphs} \label{s:dbnotation}

Given a collection of strings $\S = s_1, \ldots, s_d$ over $\A$, we prepend to each string $s_i$ $k$ copies of a symbol $\$ \notin \A$ which is lexicographically smaller than any other symbol. 
The order-$k$ {\it \dbG graph} $G(V,E)$ for the collection $\S$ is a directed edge-labeled graph containing a node $v$ for every {\bf unique $\b{k}$-mer} appearing in one of the strings of $\S$.
For each node $v$ we denote by $\repr{v} = v[1,k]$ its associated $k$-mer, where $v[1]\cdots v[k]$ are symbols. 

The graph $G$ contains an edge $(u,v)$, with label $v[k]$, iff one of the strings in $\S$ contains a {\bf $\b{(k+1)}$-mer} with
prefix $\repr{u}$ and suffix $\repr{v}$.
The edge $(u,v)$ therefore represents the $(k+1)$-mer $u[1,k] v[k]$. Note that each node has at most $\Asize$ outgoing edges and all edges incoming to node $v$ have label $v[k]$.

\begin{figure}[t]
    \centering
    \includegraphics[width=0.8\textwidth]{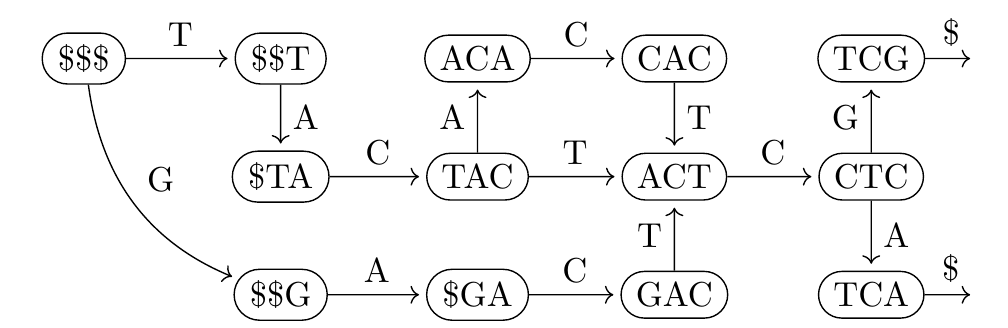}
    \caption{\dbG graph for $\S=\{$TACACT,
TACTCG, GACTCA$\}$.}
    \label{f:dbg}
\end{figure}

\begin{figure}[t]
    \centering
    \includegraphics[width=0.45\textwidth]{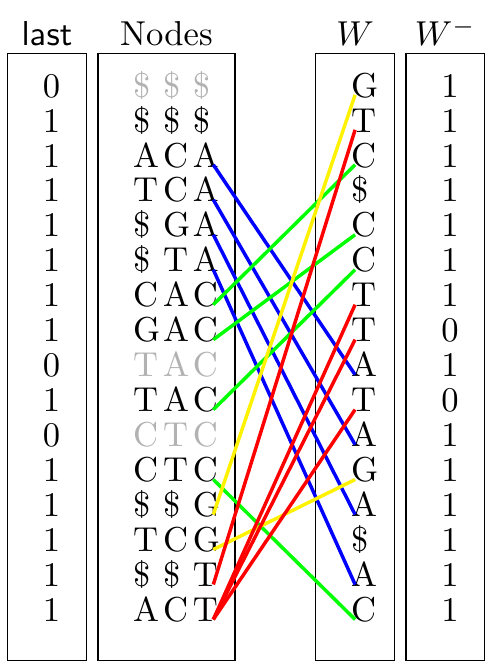}
    \caption{BOSS representation of the graph in Fig.~\ref{f:dbg}. The colored lines connect each label in $W$ to its destination node; edges of the same color have the same label. Note that edges with the same label/color reach distinct ranges of nodes, and that edges with the same label/color do not cross.}\label{f:boss}
\end{figure}

In 2012, Bowe {\em et al.}~\cite{wabi/BoweOSS12} introduced a succinct representation for the {de Bruijn} graph, usually referred to as {\em BOSS representation}, for the authors’ initials. The authors showed how to represent the graph in small space supporting fast navigation operations. The BOSS representation of the graph $G(V,E)$ is defined by considering the set of nodes $v_1, v_2, \ldots, v_n$ sorted according to the colexicographic order of their associated {$k$-mer}. 
Hence, if $\reprr{v}=v[k]\cdots v[1]$ denotes the string $\repr{v}$ reversed, the nodes are ordered so that 
\begin{equation}\label{eq:order}
\reprr{v_1} \prec \reprr{v_2} \prec \cdots \prec  \reprr{v_n}
\end{equation}
By construction the first node is $\reprr{v_1}=\$^k$ and all $\reprr{v_i}$ are distinct. 
For each node $v_i$, $i=1,\ldots,n$, we define $W_i$ as the sorted sequence of symbols on the edges leaving from node $v_i$; if $v_i$ has out-degree zero we set $W_i = \$$.  Finally, we define (see examples in Figs.~\ref{f:dbg} and~\ref{f:boss}):
\begin{enumerate}
  \item $\Wa[1,m]$ as the concatenation $W_1 W_2 \cdots W_n$;
  \item $\Wx[1,m]$ as the bitvector such that $\Wx[i]=\oneb$ iff $\Wa[i]$ corresponds to the label of the edge $(u,v)$ such that $\reprr{u}$ has the smallest rank among the nodes that have an edge going to node $v$;
  \item $\last[1,m]$ as the bitvector such that $\last[i]=1$ iff $i=m$ or the outgoing edges corresponding to $\Wa[i]$ and $\Wa[i+1]$ have different source nodes.
\end{enumerate}

The length $m$ of the arrays $\Wa$, $\Wx$, and $\last$ is equal to the number of edges plus the number of nodes with out-degree 0. In addition, the number of $\onex$'s in $\last$ is equal to the number of nodes $n$, and the number of $\onex$'s in $\Wx$ is equal to the number of nodes with positive in-degree, which is $n-1$ since $v_1=\$^k$ is the only node with in-degree 0.
Bowe {\em et al.} observed that there is a natural one-to-one correspondence, called  $\LF$ for historical reasons, between the indices $i$ such that $\Wx[i]=\onex$ and the set $\{2, \ldots,n\}$: in this correspondence $\LF(i)=j$ iff $v_j$ is the destination node of the edge associated to $\Wa[i]$. The $\LF$ correspondence is order preserving in the sense that if $\Wx[i]=\Wx[j]=\onex$ then
\begin{equation}\label{eq:wg}
\begin{array}{rcl}
\Wa[i] < \Wa[j]\; & \Longrightarrow\; &  \LF(i) < \LF(j),\\
(\Wa[i] = \Wa[j])\; \land (i < j)\; & \Longrightarrow\; & \LF(i) < \LF(j).
\end{array}
\end{equation}

Bowe {\em et al.} have shown that enriching the arrays $\Wa$, $\Wx$, and $\last$ with the data structures from~\cite{FMMN04jou,RamanRS07} supporting constant time rank and select operations, we can efficiently navigate the \dbG graph $G$.
The overall cost of encoding the three arrays and the auxiliary data structures is bounded by $m\log\sigma + 2m + \sigma \log n + o(m)$ bits, with the usual time/space tradeoffs available for rank/select data structures (see~\cite{wabi/BoweOSS12} for details).

\subsection{\WGs} \label{s:wnotation}

\begin{definition} \label{def:wg} 
A directed labeled graph $G=(V,E)$ is a {\em \WG} if there is an ordering of the nodes such that nodes with in-degree 0 precede those with positive in-degree and, for
any pair of edges \(e = (u, v)\) and \(e' = (u', v')\) labeled with $a$ and $a'$
respectively, the following monotonicity properties hold:
\begin{subequations} 
\begin{align}
a < a' & \Longrightarrow  v < v'\,,\label{eq:wgA}\\
(a = a') \land (u < u') & \Longrightarrow v \leq v'\,.\label{eq:wgB}
\end{align}
\end{subequations}\qed
\end{definition}

{It is easy to see that a \dbG graph with the nodes sorted according to~\eqref{eq:order} is a \WG.}
Informally, for the graph of Figs.~\ref{f:dbg} and~\ref{f:boss} property~\eqref{eq:wgA} follows from the fact that edges with {different labels/colors reach non interleaving} ranges of nodes, and that edges with the same label/color do not cross. Note that property~\eqref{eq:order} coincides with~\eqref{eq:wgA} and~\eqref{eq:wgB} restricted to the case in which the destination nodes are distinct. 

\begin{figure}[t]
    \centering
    \includegraphics[width=0.8\textwidth]{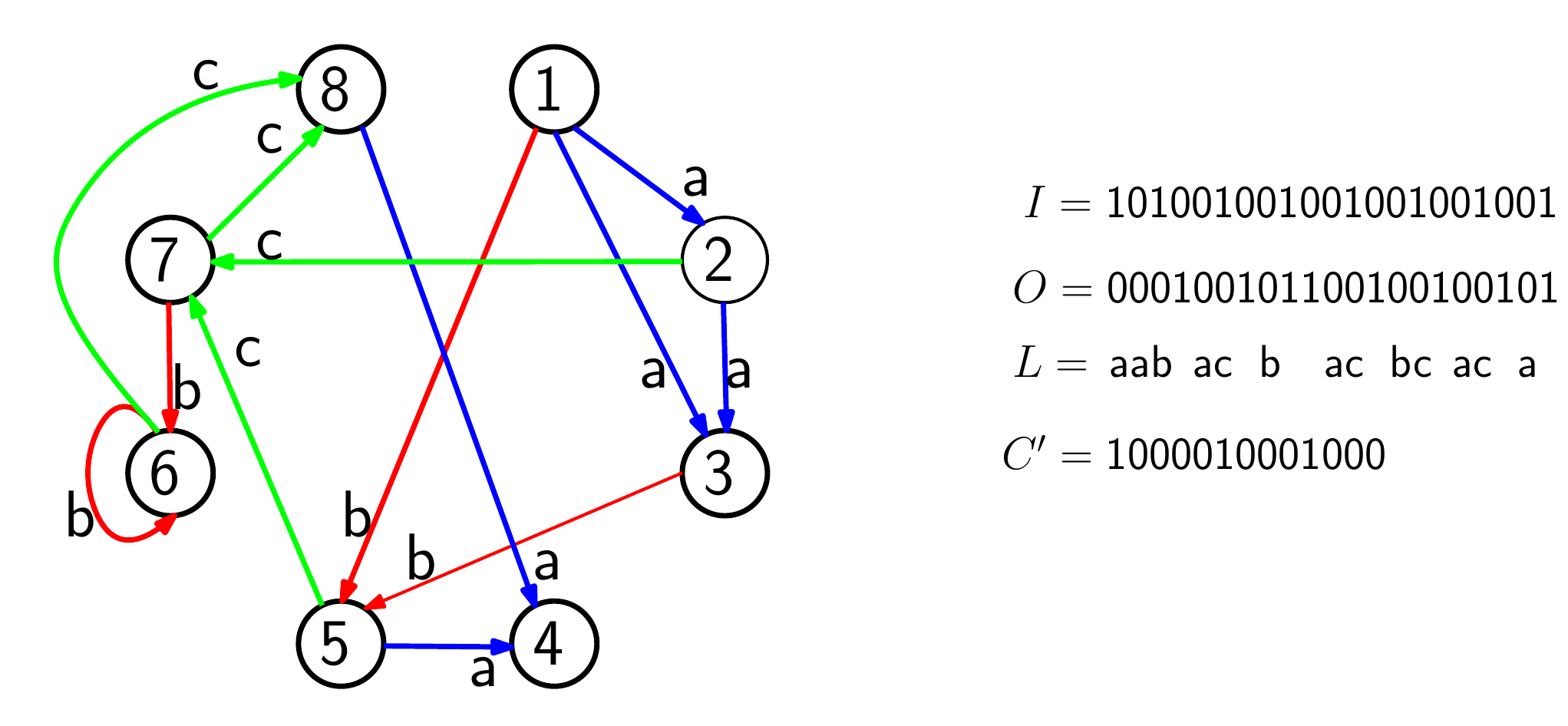}
    \caption{An example of an eight node \WG (left) and its succinct representation (right). Node 1 has no incoming edges, nodes 2--4 have incoming edges labeled {\sf a}, nodes 5--6 have incoming edges labeled {\sf b}, nodes 7--8 have incoming edges labeled {\sf c}. The binary arrays $I$ and $O$ are the unary representation of the nodes' in-degrees and out-degrees, the array $L$ stores the labels of the outgoing edges. The binary array $\Countx$ encodes the number of occurrences of any give symbol.}
    \label{fig:wg}
\end{figure}

Gagie {\em et al.}~\cite{tcs/GagieMS17} proposed the following compact representation for a \WG\ $G=(V,E)$ with $|V|=n$ and $|E|=m$. Let $x_1 < x_2 < \cdots < x_n$ denote the ordered set of nodes. For $i=1,\ldots,n$ let $\ell_i$ and $k_i$ denote respectively the out-degree and in-degree of node $x_i$. Define the binary arrays of length $n + m$
\begin{equation}\label{eq:inout}
\outx \;=\; \zerox^{\ell_1} \onex \, \zerox^{\ell_2} \onex \cdots \zerox^{\ell_n} \onex,\qquad
\inx \;=\; \zerox^{k_1}\onex \, \zerox^{k_2}\onex \cdots \zerox^{k_n}\onex.
\end{equation}
Note that $\outx$ (resp.~$\inx$) consists of the concatenated unary representations of the out-degrees (resp.~in-degrees). Let $L_i$ denote the sorted set of symbols on the edges leaving from $x_i$, and let $L[1..m]$ denote the concatenation $L = L_1 L_2 \cdots L_n$. Finally, let $\Count[1..\sigma]$ denote the array such that $\Count[c]$ is the number of edges with label smaller than $c \in \Sigma$ (we assume every distinct symbol labels some edge). As an alternative to $\Count[1..\sigma]$ one can use the binary array $\Countx[1..m]$ such that $\Countx[i]=\onex$ iff $i=1+\Count[c]$ for some $c=1..\sigma$. Fig.~\ref{fig:wg} shows an example of a \WG and its succinct representation; note that $\Count$ and $\Countx$ contains the same information as $L$, and are used only to speed up navigation.
In~\cite{tcs/GagieMS17} it is shown that we can efficiently navigate the Wheeler graph~$G$ using the arrays $\inx, \outx, L$ and $\Count$ (or $\Countx$) and auxiliary data structures supporting constant time rank/select operations.

As an example of how navigation works, suppose that given node $v$ we want to compute the smallest $u$ such that $E$ contains the edge $(u,v)$, assuming $v$ has positive indegree. 
Both $u$ and $v$ are identified by their lexicographic rank.
By construction, the desired edge corresponds to the first \zerob\ in the unary representation of $v$'s indegree. Such \zerob\ is in position \mbox{$k=1+\sel_1(\inx,v-1)$} of~$\inx$. As a running example, consider {$v=5$} in Fig.~\ref{fig:wg}; we have ${k=1+\sel_1(\inx,5-1)=10}$. The symbol on the edge is $c=\rank_1(\Countx,\rank_0(\inx,k))$, since edges in $\Countx$ are ordered by their labels in increasing order. In our running example, {$c=\rank_1(\Countx,\rank_0(\inx,10))=2$}, that is, the {second} symbol in $\A$, which is  {$\mathbf{b}$}.
The edge is the $j$-th one with label $c$ where $j=1+\rank_0(\inx,k)-\sel_1(\Countx,c)$ (there are $\rank_0(I,k)$ edges before that one, but $\sel_1(\Countx,c)$ have different labels). Hence, the symbol of that edge is the one in position $h=\sel_c(L,j)$ in $L$ and therefore corresponds to the $h$-th \zerox\ in $\outx$. In our running example, the $j$-th edge labeled with {$c = \mathbf{b}$} is given by $j=1+\rank_0(\inx,{10})-\sel_1(\Countx,{2})=1+6-6=1$, that is, $(u,v)$ is the first one with label {$\mathbf{b}$}, and {$h=\sel_b(L,1)=3$}. 
The desired node $u$ is therefore the node whose outdegree unary representation in $\outx$ contains the $h$-th bit {\zerox} of $\outx$, that is $u = 1+\rank_1(\outx,\sel_0(\outx,h))$. This, in our running example, is {$u=1+\rank_1(\outx,\sel_0(\outx,3))=1+\rank_1(\outx,3)=1+0=1$}.
Therefore, the edge {$(u,v)=(1,5)$} is labeled with {$\mathbf{b}$}.

More in general, combining~\cite{tcs/GagieMS17} with the succinct representations from~\cite{BNtalg14} we get the following result. 

\begin{lemma}\label{lemma:WGspace}
It is possible to represent an $n$-node, $m$-edge \WG\ with
labels over the alphabet $\A$ in $2(n+m) + m\log\sigma +\sigma \log m + o(n + m\log\sigma)$ bits. The representation supports forward and backward traversing of the edges in $\Oh(1)$ time assuming $\sigma=w^{\Oh(1)}$, where $w$ is the word size.\qed
\end{lemma}

Note that there are strong similarities between the BOSS representation and the \WG succinct representation, with the arrays $\Wa$ and $\last$ corresponding respectively to $L$, and $\outx$, 
while the array $\Wx$ is a permutation of $\inx$.
This is not surprising since the latter was inspired by the former. The BOSS representation is more efficient since we can assume each (in)out-degree $d$ is positive so we unary encode $d-1$ instead of $d$ saving $n$ bits in $\Wx$ and $\last$. (For simplicity we omitted from the BOSS representation an array, called $F$ in~\cite{wabi/BoweOSS12}, which corresponds to the $\Count$ array of \WGs). 

\begin{figure}
    \centering
    \includegraphics[width=0.8\textwidth]{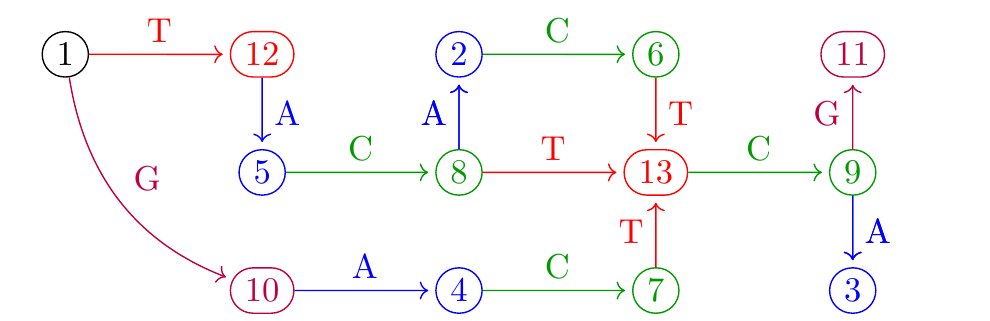}\\
    \setlength{\tabcolsep}{2pt}
    \begin{tabular}{rl}
              & \\
        $I =$ & \texttt{101010101010101010101010001} \\
        $O =$ & \texttt{001011010101010010010110101}\\
        $L =$ & \texttt{TG C~~C C T T AT AG A~~A~C}\\
        $C'=$ & \texttt{10001000101000}
    \end{tabular}
    \caption{\WG representation for the \dbG graph presented in Fig.~\ref{f:dbg}.}
    \label{f:dbg-wg}
\end{figure}

Fig.~\ref{f:dbg-wg} shows the \dbg graph of Fig.~\ref{f:dbg} as a \WG and its corresponding succinct representation. Note that in the \WG representation the outgoing edges labeled \$ from nodes 3 and 11 in are not necessary since, since the representation allows edges with out-degree zero.

The importance of Wheeler graphs comes from the observation in~\cite{tcs/GagieMS17} that many succinct data structures supporting efficient substring queries~\cite{Durbin2014efficient,FM05,jacm09,talg/FerraginaV10,tcs/MantaciRRS07,Na2018,Siren2017} can be seen as Nondeterministic Finite Automata (NFA) whose states can be sorted so that the resulting graph is a \WG. We leave the details to~\cite{tcs/GagieMS17} and formalize this notion with the following definitions.

\begin{definition} \label{def:nfa} 
A nondeterministic finite automata (NFA) is a quintuple \mbox{$\mathcal{A}=(V, E, F, s, \Sigma)$}, where $V$ is a set of states (or nodes), $\Sigma$ is the alphabet (set of labels),
$E \subseteq V \times V \times \Sigma$ is a set of directed labeled edges, $F\subseteq V$ is a set of accepting states, and $s\in V$ is the start state. We additionally require that $s$ is the only state with in-degree 0, that each node is reachable from~$s$, and that from each node we can reach a final state.\qed 
\end{definition}

\begin{definition}\label{def:wa}
A {\em \WA} is a NFA without $\epsilon$-transitions for which there is an ordering of the states that makes the state diagram a Wheeler graph.\qed
\end{definition}

The reader should refer to~\cite{Alanko2020} to see that Definition~\ref{def:nfa} is not restrictive and for further details and properties of \WAs. With a little abuse of notation in the following we will use the terms \WG and \WA as synonymous.

\section{Merging BOSS representations of \dbG graphs}\label{s:alg}

Suppose we are given the BOSS {representations} of two \dbG graphs $\langle \Waz, \Wxz, \lastz \rangle$ and $\langle \Wao, \Wxo, \lasto \rangle$ obtained respectively from the collections of strings $\Colz$ and $\Colo$. 
In this section we show how to compute the BOSS representation for the union collection $\Colzo = \Colz \cup \Colo$. 
The procedure does not change in the general case when we are merging an arbitrary number of graphs. \added{In Section~\ref{sec:stateoftheart} we compare our solution with the state of the art algorithm for the same problem.}

Let $\Gz$ and $\Go$ denote respectively the (uncompressed) \dbG graphs for $\Colz$ and $\Colo$, and let 
$$
v_1, \ldots, v_{\nz}\qquad\mbox{and}\qquad w_1, \ldots, w_{\none}
$$
denote their respective set of nodes sorted in colexicographic order. Hence, with the notation of the previous section we have
\begin{equation}\label{eq:sorted}
\reprr{v_1} \prec \cdots \prec \reprr{v_\nz} \qquad\mbox{and}\qquad 
\reprr{w_1} \prec \cdots \prec \reprr{w_\none}
\end{equation}
We observe that the $k$-mers in the collection $\Colzo$ are simply the union of the $k$-mers in $\Colz$ and $\Colo$. To build the \dbG graph for $\Colzo$ we need therefore to: 1) merge the nodes in $\Gz$ and $\Go$ according to the colexicographic order of their associated $k$-mers, 2) recognize when two nodes in $\Gz$ and $\Go$ refer to the same $k$-mer, and 3) properly merge and update the bitvectors $\Wxz$, $\lastz$ and $\Wxo$, $\lasto$.

\subsection{Phase 1: Merging $k$-mers} \label{ss:phase1}

The main technical difficulty is that in the BOSS representation the $k$-mers associated to each node $\repr{v}=v[1,k]$ are not directly available. 
Our algorithm will reconstruct them using the symbols associated to the graph edges; to this end the algorithm will consider only the edges such that the corresponding {entries} in $\Wxz$ or $\Wxo$ {are} equal to $\onex$. 
Following these edges, first we recover the last symbol of each $k$-mer, following them a second time we recover the last two symbols of each $k$-mer and so on. However, to save space we do not explicitly maintain the $k$-mers; instead, using the ideas from~\cite{bcb/HoltM14,bioinformatics/HoltM14} our algorithm computes a bitvector $\bv{k}$ representing how the $k$-mers in $\Gz$ and $\Go$ should be merged according to the colexicographic order.  

To this end, our algorithm executes $k-1$ iterations of the code shown in Fig.~\ref{fig:xHMalgo} (note that lines~\ref{line:B=0h}--\ref{line:B=0hEnd} and~\ref{line:block_process_start}--\ref{line:block_process_end} of the algorithm are related to the computation of the $B$ array that is used in the following section). For $h=2,3,\ldots,k$, during iteration $h$, we compute the bitvector $\bv{h}[1,n_0+n_1]$ containing $n_0$ \zerob's and $n_1$ \oneb's such that $\bv{h}$ satisfies the following property

\begin{property}\label{prop:xhblock}
For $i=1,\ldots, \nz$ and $j=1,\ldots \none$ the $i$-th \zerob\ precedes the
$j$-th \oneb\ in $\bv{h}$ if and only if 
$\reprr{v_i}[1,h] \;\preceq\; \reprr{w_j}[1,h]$.
\qed
\end{property}

Property~\ref{prop:xhblock} states that if we merge the nodes from $\Gz$ and $\Go$ according to the bitvector $\bv{h}$ the corresponding $k$-mers will be sorted according to the lexicographic order restricted to the first $h$ symbols of each reversed $k$-mer. As a consequence, $\bv{k}$ will provide us the colexicographic order of all the nodes in $\Gz$ and $\Go$. 
To prove that Property~\ref{prop:xhblock} holds, we first define $\bv{1}$ and show that it satisfies the property, then we prove that for $h=2,\ldots,k$ the code in Fig.~\ref{fig:xHMalgo} computes $\bv{h}$ that still satisfies Property~\ref{prop:xhblock}. 

For $c\in\A$ let $\laz{c}$ and $\lao{c}$ denote respectively the number of nodes in $\Gz$ and $\Go$ whose associated $k$-mers end with symbol~$c$. These values can be computed with a single scan of $\Waz$ (resp. $\Wao$) considering only the symbols $\Waz[i]$ (resp. $\Wao[i]$) such that $\Wxz[i]=\onex$ (resp. $\Wxo[i]=\onex$). By construction, it is
$$
\nz = 1 + \sum_{c\in\A} \laz{c},\qquad\mbox\qquad \none = 1 + \sum_{c\in\A} \lao{c}
$$
where the two 1's account for the nodes $v_1$ and $w_1$ whose associated $k$-mer is $\xx^k$. 
We define 
\begin{equation}\label{eq:Z1}
\bv{1} = \u{\zerob\oneb}~ \u{\zerob^{\laz{1}}\oneb^{\lao{1}}}~ \u{\zerob^{\laz{2}}\oneb^{\lao{2}}} \cdots \u{\zerob^{\laz{\sigma}}\oneb^{\lao{\sigma}}}\;.
\end{equation}
The first pair \zerob\oneb\ in $\bv{1}$ accounts for $v_1$ and $w_1$; for each $c\in\A$ the group $\zerob^{\laz{c}}\oneb^{\lao{c}}$ accounts for the nodes ending with symbol~$c$. Note that, apart from the first two symbols, $\bv{1}$ can be logically partitioned into $\sigma$ subarrays one for each alphabet symbol. For $c\in\A$ let 
$$\sta(c) = 3 + \sum_{i<c}(\laz{i} + \lao{i})$$
then the subarray corresponding to $c$ starts at position $\sta(c)$ and has size $\laz{c} + \lao{c}$. 
As a consequence of~\eqref{eq:sorted}, the $i$-th \zerob\ (resp. $j$-th \oneb) belongs to the subarray associated to symbol $c$ iff $\reprr{v_i}[1]=c$ (resp. $\reprr{w_j}[1]=c$). 

\begin{figure}[t]
\hrule\smallbreak
\begin{algorithmic}[1]
\For{$c \gets 1$ \KwTo $\sigma$} \label{line:initloop}
  \State $F[c] \gets \sta(c)$    \Comment{Init $F$ array}
  \State $\Bid[c] \gets -1$      \Comment{Init $\Bid$ array}
\EndFor
\State $i_0 \gets i_1 \gets 1$  \Comment{Init counters for $\Waz$ and $\Wao$}
\State $\bv{h} \gets \zerob\oneb$  \Comment{First two entries correspond to {$v_1$ and $w_1$}}\label{step:01}
\For{$p \gets 1$ \KwTo $n_0+n_1$}\label{line:initmainloop}
    \If{$B[p]\neq \sbot $ \KwAnd $B[p]\neq h$}\label{line:B=0h}
       \State $\bid\gets p$\Comment{A new block of $\bv{h-1}$ is starting}\label{line:blockstart}
    \EndIf  \label{line:B=0hEnd}
    \State $b \gets \bv{h-1}[p]$\Comment{Get bit $b$ from $\bv{h-1}$}
    \Repeat{}\Comment{Current node is from graph $G_b$}
      \If {$\Wxx{b}[i_b]=\oneb$}\label{line:W=1}
        \State $c \gets \Wax{b}[i_b]$ \Comment{Get symbol from outgoing edges}\label{step:getc}
        \State $q \gets F[c]{\mathsf ++}$ \Comment{Get destination for $b$ according to symbol $c$}
        \State $\bv{h}[q] \gets b$        \Comment{Copy bit $b$ to $\bv{h}$}\label{line:updatebv}\label{step:putc}
         \If{$\Bid[c]\neq \bid$}\label{line:block_process_start}
          \State $\Bid[c]\gets \bid$\Comment{Update block id for symbol $c$}
          \If{$B[q] = \sbot$} \Comment{Check if already marked}\label{line:new_start}
            \State$B[q] \gets h$\Comment{A new block of $\bv{h}$ will start here}\label{line:writeh}
          \EndIf\label{line:writehEndIf}
         \EndIf \label{line:block_process_end}        
      \EndIf 
    \Until{$\lastx{b}[i_b{\mathsf ++}] \neq \onex$} \Comment{Exit if $c$ was last edge}
\EndFor\label{line:endmainloop}
\end{algorithmic}
\smallbreak\hrule
\caption{Main procedure for merging succinct \dbG graphs. Lines~\ref{line:B=0h}--\ref{line:B=0hEnd} and~\ref{line:block_process_start}--\ref{line:block_process_end} are related to the computation of the $B$ array introduced in Section~\ref{ss:phase2}.}\label{fig:xHMalgo}
\end{figure}

To see that $\bv{1}$ satisfies Property~\ref{prop:xhblock}, observe that the $i$-th \zerob\ precedes $j$-th \oneb\ iff the $i$-th \zerob\ belongs to a subarray corresponding to a symbol not larger than the symbol corresponding to the subarray containing the $j$-th \oneb; this implies $\reprr{v_i}[1,1] \preceq \reprr{w_j}[1,1]$. 

The bitvectors $\bv{h}$ computed by the algorithm in Fig.~\ref{fig:xHMalgo} can be logically divided into the same subarrays we defined for $\bv{1}$. In the algorithm we use an {array $F[1,\sigma]$ to keep track of the next available position of each subarray}. Because of how the array $F$ is initialized and updated, we see that every time we read a symbol $c$ at line~\ref{step:getc} the corresponding bit $b=\bv{h-1}[k]$, which denotes the graph $\Gb$ containing~$c$, is written in the portion of $\bv{h}$ corresponding to $c$ (line~\ref{step:putc}). The only exception are the first two entries of $\bv{h}$ which are written at line~\ref{step:01} which corresponds to the nodes $v_1$ and $w_1$. We treat these nodes differently since they are the only ones with in-degree zero. For all other nodes, we implicitly use the one-to-one correspondence {\eqref{eq:wg}} between entries $W[i]$  with $\Wx[i]=\oneb$ and nodes $v_j$ with positive in-degree. 

The following Lemma proves the correctness of the algorithm in Fig.~\ref{fig:xHMalgo}.

\begin{lemma}\label{lemma:xhblock}
For $h=2,\ldots,k$, the array $\bv{h}$ computed by the algorithm in Fig.~\ref{fig:xHMalgo} satisfies Property~\ref{prop:xhblock}.
\end{lemma}

\begin{proof}
To prove the ``if'' part of Property~\ref{prop:xhblock} let $1 \leq f < g \leq \nz+\none$ denote two indexes such that {$\bv{h}[f]$} is the $i$-th \zerob\ and {$\bv{h}[g]$} is the $j$-th \oneb\ in $\bv{h}$ for some $1 \leq i \leq \nz$ and $1 \leq j \leq \none$. We need to show that $\reprr{v_i}[1,h] \preceq \reprr{w_j}[1,h]$. 

Assume first $\reprr{v_i}[1]\neq \reprr{w_j}[1]$. The hypothesis $f<g$ implies $\reprr{v_i}[1]<\reprr{w_j}[1]$, since otherwise during iteration~$h$ the $j$-th \oneb\ would have been written in a subarray of $\bv{h}$ preceding the one where the $i$-th \zerob\ is written. Hence $\reprr{v_i}[1,h] \preceq \reprr{w_j}[1,h]$ as claimed. 
 
Assume now $\reprr{v_i}[1] = \reprr{w_j}[1] = c$. In this case during iteration $h$
the $i$-th \zerob\ and the $j$-th \oneb\ are both written to the subarray of $\bv{h}$ associated to symbol~$c$. Let $f'$, $g'$ denote respectively the value of the main loop variable~$p$ in the procedure of Fig.~\ref{fig:xHMalgo} when the entries {$\bv{h}[f]$} and {$\bv{h}[g]$} are written. Since each subarray in $\bv{h}$ is filled sequentially, the hypothesis $f<g$ implies $f'<g'$. By construction $\bv{h-1}[{f}']=\zerox$ and $\bv{h-1}[{g}']=\onex$. Say ${f}'$ is the $i'$-th \zerob\ in $\bv{h-1}$ and ${g}'$ is the $j'$-th \oneb\ in $\bv{h-1}$. By the inductive hypothesis on $\bv{h-1}$ it is
\begin{equation}\label{eq:xhblock2}
\reprr{v_{i'}}[1,h-1] \;\preceq\; \reprr{w_{j'}}[1,h-1].
\end{equation}
By construction there is an edge labeled $c$ from $v_{i'}$ to $v_i$ and from $w_{j'}$ to $w_j$ hence 
$$
\repr{v_{i}}[k-h,k] = \repr{v_{i'}}[k-h+1,k]c,\qquad \repr{w_{j}}[k-h,k] = \repr{w_{j'}}[k-h+1,k]c;
$$
therefore
$$
\reprr{v_{i}}[1,h] = c \reprr{v_{i'}}[1,h-1],\qquad \reprr{w_{j}}[1,h] = c \reprr{w_{j'}}[1,h-1];
$$
using~\eqref{eq:xhblock2} we conclude that $\reprr{v_i}[1,h] \preceq \reprr{w_j}[1,h]$ as claimed.

For the ``only if'' part of Property~\ref{prop:xhblock}, assume $\reprr{v_i}[1,h] \preceq \reprr{w_j}[1,h]$ for some $i\geq 1$ and $j\geq 1$. We need to prove that in $\bv{h}$ the $i$-th \zerob\ precedes the $j$-th \oneb. If $\reprr{v_i}[1]\neq\reprr{w_j}[1]$ the proof is immediate. If $c=\reprr{v_i}[1]=\reprr{w_j}[1]$ then
$$
\reprr{v_i}[2,h]\preceq\reprr{w_j}[2,h].
$$
Let $i'$ and $j'$ be such that $\reprr{v_{i'}}[1,h-1] = \reprr{v_i}[2,h]$ and $\reprr{w_{j'}}[1,h-1] =\reprr{w_j}[2,h]$.
By induction {hypothesis}, in $\bv{h-1}$ the $i'$-th \zerob\ precedes the $j'$-th \oneb. 

During phase~$h$, the $i$-th \zerob\ in $\bv{h}$ is written to position $f$ when processing the $i'$-th \zerob\ of $\bv{h-1}$, and the $j$-th \oneb\ in $\bv{h}$ is written to position $g$ when processing the $j'$-th \oneb\ of $\bv{h-1}$. Since in $\bv{h-1}$ the $i'$-th \zerob\ precedes the $j'$-th \oneb\ and since $f$ and $g$ both belong to the subarray of $\bv{h}$ corresponding to the symbol $c$, their relative order does not change and the $i$-th \zerob\ precedes the $j$-th \oneb\ as claimed.\qed
\end{proof}

\subsection{Phase 2: Recognizing identical $k$-mers} \label{ss:phase2}

Once we have determined, via the bitvector $\bv{h}[1, n_0+n_1]$, the colexicographic order of the $k$-mers, we need to determine when two $k$-mers are identical since in this case we have to merge their outgoing and incoming edges. Note that two identical $k$-mers will be consecutive in the colexicographic order and they will necessarily belong one to $\Gz$ and the other to $\Go$.

Following Property~\ref{prop:xhblock}, we identify the $i$-th \zerob\ in
$\bv{h}$ with $\reprr{v_i}$ and the $j$-th \oneb\ in $\bv{h}$ with $\reprr{w_j}$.
\added{For $h=2,\dots,k$, let $\kh+1$ be the number of $h$-blocks.}
Property~\ref{prop:xhblock} is equivalent to state that we can logically
partition $\bv{h}$ into $\kh+1$ $h$-blocks
\begin{equation}\label{eq:Zblocks}
\bv{h}[1,\ell_1],\; \bv{h}[\ell_1+1, \ell_2],\; \ldots,\;
\bv{h}[\ell_\kh+1,n_0+n_1]
\end{equation}
such that each block corresponds to a set of $k$-mers which are prefixed by the same length-$h$ substring.
Note that during iterations {$h=2,3,\dots,k$} the $k$-mers within an $h$-block will be rearranged, and sorted according to longer and longer prefixes, but they will stay within the same block. 

In the algorithm of Fig.~\ref{fig:xHMalgo}, in addition to $\bv{h}$, we maintain an integer array $B[1,\nz+\none]$, such that at the end of iteration~$h$ it is $B[i]\neq 0$ if and only if a block of $\bv{h}$ starts at position~$i$. Initially, for $h=1$, since we have one block {per} symbol, we set
$$
B=\u{1 0}\, \u{1 0^{\laz{1}+\lao{1}-1}}\, \u{1 0^{\laz{2}+\lao{2}-1}} \cdots \u{10^{\laz{\sigma}+\lao{\sigma}-1}}.
$$
During iteration~$h$, new block boundaries are established as follows. At line~\ref{line:blockstart} we identify each existing block with its starting position. Then, at lines~\ref{line:block_process_start}--\ref{line:block_process_end}, if the entry $\bv{h}[q]$ corresponds to a $k$-mer that has the form $c\alpha$, while $\bv{h}[q-1]$ to one with form $c\beta$, with $\alpha$ and $\beta$ belonging to different blocks, then we know that $q$ is the starting position of an $h$-block. Note that we write $h$ to $B[q]$ only if no other value has been previously written there. This ensures that $B[q]$ is the smallest position in which the strings corresponding to $\bv{h}[q-1]$ and $\bv{h}[q]$ differ, or equivalently, $B[q]-1$ is the LCP between the strings corresponding to $\bv{h}[q-1]$ and $\bv{h}[q]$. The above observations are summarized in the following Lemma, which is a generalization to \dbG\ graphs of an analogous result for BWT merging established in Corollary~4 in~\cite{spire/EgidiM17}.

\begin{lemma}\label{lemma:lcp}
After iteration~$k$ of the merging algorithm for $q=2,\ldots, \nz+\none$ if $B[q]\neq 0$ then $B[q]-1$ is the LCP between the {reverse} $k$-mers corresponding to $\bv{k}[q-1]$ and $\bv{k}[q]$, while if $B[q]=0$ {their LCP is equal to $k$}, hence such $k$-mers are equal.\qed
\end{lemma}

The above lemma shows that using array $B$ we can establish when two $k$-mers are equal and consequently the associated graph nodes should be merged. 

\subsection{Phase 3: Building BOSS representation for the union graph}
We now show how to compute the succinct representation of the union graph $\Gz~\cup~\Go$, consisting of the arrays $\langle \Wax{01}$, $\Wxx{01}$, $\lastx{01}\rangle$, given the succinct representations of $\Gz$ and $\Go$ and the arrays $\bv{k}$ and $B$.

The arrays  $\Wax{01}$, $\Wxx{01}$, $\lastx{01}$ are initially empty and we fill them in a single {sequential} pass. 
For $q=1,\ldots,\nz+\none$ we consider the values $\bv{k}[q]$ and $B[q]$.
If $B[q]=0$ then the $k$-mer associated to $\bv{k}[q-1]$, say $\reprr{v_i}$ is identical to the $k$-mer associated to  $\bv{k}[q]$, say $\reprr{w_j}$. In this case we recover from $\Wax{0}$ and $\Wax{1}$ the labels of the edges outgoing from $v_i$ and $w_j$, we compute their union and write them to $\Wax{01}$ {(we assume the edges are in lexicographic order)}, writing at the same time the 
representation of the out-degree of the new node to $\lastx{01}$. 
If instead $B[q]\neq 0$, then the $k$-mer associated to $\bv{k}[q-1]$ is unique and we copy the information of its outgoing edges {and out-degree} directly to $\Wax{01}$ and $\lastx{01}$.
When we write the symbol $\Wax{01}[i]$ we simultaneously write the bit $\Wxx{01}[i]$ according to the following strategy. 
If the symbol $c=\Wax{01}[i]$ is the first occurrence of $c$ after a value $B[q]$, with $0 < B[q] < k$, then we set $\Wxx{01}[i]=\oneb$, otherwise we set $\Wxx{01}[i]=\zerob$.
The rationale is that if no values $B[q]$ with $0 < B[q] < k$  occur between two nodes, then the associated (reversed) $k$-mers have a common LCP of length $k-1$ and therefore if they both have an outgoing edge labeled with $c$ they reach the same node and only the first one  should have $\Wxx{01}[i]=\oneb$. 

\subsection{Implementation details and analysis}\label{s:implementation}

Let $n=\none+\nz$ denote the sum of number of nodes in $\Gz$ and $\Go$, and let $\edges=|\Waz|+|\Wao|$ denote the sum of the number of edges. The $k$-mer merging algorithm as described executes in $\Oh(\edges)$ time a first pass over the arrays $\Waz$, $\Wxz$, and $\Wao$, $\Wxo$ to compute the values $\laz{c} + \lao{c}$ for $c\in\A$ and initialize the arrays {$F[1,\sigma]$, $\sta[1,\sigma]$, $\Bid[1,\sigma]$ }and $\bv{1}[1,n]$ (Phase 1). 
Then, the algorithm executes $k-1$ iterations of the code in Fig.~\ref{fig:xHMalgo} each iteration taking $\Oh(\edges)$ time. Finally, still in $\Oh(\edges)$ time the algorithm computes the succinct representation of the union graph (Phases 2 and 3). The overall running time is therefore $\Oh(\edges\, k)$. 

We now analyze the space usage of the algorithm.  In addition to the input and the output, our algorithm uses $2n$ bits for two instances of the $\bv{\cdot}$ array (for the current $\bv{h}$ and for the previous $\bv{h-1}$), plus $n\lceil\log k\rceil$ bits for the $B$ array. 
Note, however, that during iteration $h$ we only need to check whether $B[i]$ is equal to 0, $h$, or some value within 0 and $h$. Similarly, for the computation of $\Wxx{01}$ we only need to distinguish between the cases where $B[i]$ is equal to 0, $k$ or some value $0 < B[i]< k$.
Therefore, we can save space replacing $B[1,n]$ with an array $\Bxx[1,n]$ containing two bits per entry representing the four possible states $\{\zzx,\oddx,\evenx,\oox\}$. During iteration $h$, the values in $\Bxx$ are used instead of the ones in $B$ as follows: An entry $\Bxx[i]=\zzx$ corresponds to $B[i]=0$, an entry $\Bxx[i]=\oox$ corresponds to an entry $0 < B[i] < h-1$. In addition, if $h$ is even, an entry $\Bxx[i]=\evenx$ corresponds to $B[i]=h$ and an entry $\Bxx[i]=\oddx$ corresponds to $B[i]=h-1$; while if $h$ is odd the correspondence is $\evenx \rightarrow h-1$, $\oddx \rightarrow h$. 
The reason for this apparently involved scheme, first introduced in~\cite{wabi/EgidiLMT18}, is that during phase $h$, an entry in $\Bxx$ can be modified either before or after we have read it at Line~\ref{line:blockstart}. \added{To update $\Bxx$ it suffices to replace Lines~\ref{line:new_start}--\ref{line:writehEndIf} with instructions for setting to \oneb\ the appropriate bit of $\Bxx[q]$. In two iterations these updates will correctly transform a value $\Bxx[i] = \zzx$, meaning $B[i]=0$, into the value $\Bxx[i]=\oox$, meaning $0<B[i]<h-1$. For instance if, when $h$ is even, $\Bxx[i]$ is set to $\evenx$, at the following iteration $h'=h+1$ (odd), $\Bxx[i]=\evenx$ will stand for $B[i]=h'-1$, and set to $\Bxx[i]=\oox$. Then at the following iterations, $h''>h'$, $\Bxx[i]=\oox$ stands for $0 < B[i] < h''-1$. } 
Using this technique, the working space of the algorithm, i.e., the space in addition to the input and the output, is $4n$ bits plus $3\sigma + \Oh(1)$ words of RAM for the arrays $\sta$, {$F$}, and $\Bid$.

\begin{theorem}\label{t:merge1}
The merging of two succinct representations of two order-$k$ \dbG\ graphs can be done in $\Oh(\edges\, k)$ time using $4n$ bits plus $\Oh(\sigma)$ words of working space.\qed
\end{theorem}

We stated the above theorem in terms of working space, since the total space depends on how we store the input and output, and for such storage there are several possible alternatives. The usual assumption is that the input \dbG graphs, i.e. the arrays $\langle \Waz, \Wxz, \lastz \rangle$ and $\langle \Wao, \Wxo, \lasto \rangle$, are stored in RAM using overall $m\log \sigma + 2m$ bits. Since the three arrays representing the output \dbG graph are generated sequentially in one pass, they are usually written directly to disk without being stored in RAM, so they do not contribute to the total space usage. Also note that during each iteration of the algorithm in Fig.~\ref{fig:xHMalgo}, the input arrays are all accessed sequentially. Thus we could keep them on disk reducing the overall RAM usage to just $4n$ bits plus $\Oh(\sigma)$ words; the resulting algorithm would perform additional $\Oh( k(m\log \sigma + 2m)/D )$ I/Os where $D$ denotes the disk page size in bits.

\subsection{Comparison with the state of the art}\label{sec:stateoftheart} The \dbG graph merging algorithm by Muggli {\em et al.}~\cite{MuggliBC19,Muggli229641} is similar to ours in that it has a {\em planning phase} consisting of the colexicographic sorting of the $(k+1)$-mers associated to the edges of $G_0$ and $G_1$. To this end, the algorithm uses a standard MSD radix sort. However only the most significant symbol of each $(k+1)$-mer is readily available in $\Waz$ and $\Wao$. Thus, during each iteration the algorithm computes also the next symbol of each $(k+1)$-mer that will be used as a sorting key in the next iteration. The overall space for such symbols is $2m\lceil \log \sigma\rceil$ bits, since for each edge we need the symbol for the current and next iteration. In addition, the algorithm {uses up to $2(n+m)$ bits}
to maintain the set of intervals consisting in edges whose associated reversed $(k+1)$-mer have a common prefix; these intervals correspond to the blocks we implicitly maintain in the array $\Bxx$ using only $2n$ bits.

Summing up, the algorithm by Muggli {\em et al.} runs in $\Oh(mk)$ time, and uses $2(m\lceil\log\sigma\rceil + m + n)$ bits plus $\Oh(\sigma)$ words of working space. Our algorithm has the same time complexity but uses less space: even for $\sigma=5$ as in bioinformatics applications, our algorithm uses less than half the space ($4n$ bits vs. $6.64 m+2n$ bits).
This space reduction significantly influences the size of the largest \dbG graph that can be built with a given amount of RAM. For example, in the setting in which the input graphs are stored on disk and all the RAM is used for the working space, our algorithm can build a \dbG graph whose size is twice the size of the largest \dbG graph that can be built with the algorithm of Muggli \etal.

We stress that the space reduction was obtained by substantially changing the sorting procedure. Although both algorithms are based on radix sorting they differ substantially in their execution. The algorithm by Muggli {\em et al.} follows the traditional MSD radix sort strategy; hence it establishes, for example, that $ACG \prec ACT$ when it compares the third `digits` and finds that $G < T$. In our algorithm we use a mixed LSD/MSD strategy: in the above example we also find that $ACG \prec ACT$ during the third iteration, but this is established without comparing directly $G$ and $T$, which are not explicitly available. Instead, during the second iteration the algorithm finds that $CG \prec CT$ and during the third iteration it uses this fact to infer that $ACG \prec ACT$: this is indeed a remarkable sorting trick first introduced in~\cite{bioinformatics/HoltM14} and adapted here to \dbG graphs.

\subsection{Merging colored and variable order representations}\label{s:variants}

The {\em colored \dbG graph}~\cite{iqbal2012a} is an extension of \dbG graphs for a collection of graphs, where each edge is associated with a set of ``colors'' that indicates which graphs contain that edge. The BOSS representation for a set of graphs $\mathcal{G} = \{G_1, \dots, G_t\}$ contains the union of all individual graphs. In its simplest representation, 
the colors of all edges $W[i]$ are stored in a two-dimensional binary array $\mathcal{M}$, such that $\mathcal{M}[i,j]=1$ iff {the} $i$-th edge is present in graph $G_j$.
There are different compression alternatives for the color matrix $\mathcal{M}$ that support fast operations~\cite{wabi/AlmodaresiPP17,bioinformatics/MarcusLS14,bioinformatics/MuggliBNMBRGPB17}.
Recently, Alipanah {\em et~al.}~\cite{spire/AlipanahiKB18} presented a different approach to reduce the size of $\mathcal{M}$ by recoloring.

Another variant of \dbg graph is the {\em variable order succinct \dbg graph}. The order $k$ of a \dbg graph is an important parameter for genome assembling algorithms. 
When $k$ is small the graph can be too small and uninformative, whereas when $k$ is large the graph can become too large or disconnected. To add flexibility to the BOSS representation, Boucher {\em et al.}~\cite{dcc/BoucherBGPS15} suggest to enrich the BOSS representation of an order-$k$ \dbg graph with the length of the longest common suffix ($\LCS$) between the $k$-mers of consecutive nodes $v_1, v_2, \dots, v_n$ sorted according to~\eqref{eq:order}. These lengths are stored in a wavelet tree using $O(n \log k)$ additional bits. The authors show that this enriched representation supports navigation on all \dbG\ graphs of order $k'\leq k$ and that it is even possible to vary the order $k'$ of the graph on the fly during the navigation up to the maximum value $k$. The $\LCS$ between $\repr{v_i}$ and $\repr{v_{i+1}}$ is equivalent to the length of the longest common prefix ($\LCP$) between their reverses $\reprr{v_i}$ and $\reprr{v_{i+1}}$. The $\LCP$ (or $\LCS$) between the nodes $v_1, v_2,  \cdots, v_n$ can be computed during the $k$-mer sorting phase. In the following we denote by VO-BOSS the {variable order succinct \dbG graph} consisting of the BOSS representations enriched with the $\LCS/\LCP$ information.

In this section we show that our algorithm can be easily generalized to merge colored and VO-BOSS representations. {Note that the merging algorithm by Muggli {\em et al.} can also merge colored BOSS representations, but in its original formulation, it cannot merge VO-BOSS representations.}

Given the colored BOSS representation of two \dbG graphs $\Gz$ and $\Go$, the corresponding color matrices $\mathcal{M}_0$ and $\mathcal{M}_1$ have size $m_0 \times c_0$ and $m_1 \times c_1$. 
We initially create a new color matrix $\mathcal{M}_{01}$ of size $(m_0+m_1) \times (c_0+c_1)$ with all entries empty.
During the merging {of the union graph (Phase 3)}, for $q=1,\ldots,n$, we write the colors of the edges associated to $\bv{h}[q]$ to the corresponding line in $\mathcal{M}_{01}$  {possibly merging the colors when we find nodes with identical $k$-mers}  {in $\Oh(c_{01})$ time, with $c_{01}=c_0+c_1$.} 
To make sure that color {\sf id}'s from $\mathcal{M}_{0}$ are different from those in $\mathcal{M}_{1}$  in the new graph we add the constant $c_0$ (the number of distinct colors in $\Gz$) to any color {\sf id} coming from the matrix $\mathcal{M}_1$.  

\begin{theorem}
The merging of two succinct representations of colored \dbG\ graphs takes $\Oh(\edges \, \max(k,c_{01}))$ time and $4n$ bits  plus $\Oh(\sigma)$ words of working space, where $c_{01} = c_0+c_1$. \qed
\end{theorem}

We now show that we can compute the variable order VO-BOSS representation of the union of two \dbG graphs $G_0$ and $G_1$ given their {\em plain}, eg. non variable order, BOSS representations. 
For the VO-BOSS representation we need the $\LCS$ array for the nodes in the union graph $\langle \Wax{01}$, $\Wxx{01}$, $\lastx{01}\rangle$.
Notice that after merging the $k$-mers of $\Gz$ and $\Go$ with the algorithm in Fig.~\ref{fig:xHMalgo} (Phase~1) the values in $B[1,n]$ already provide the LCP information between the reverse labels of all consecutive nodes (Lemma~\ref{lemma:lcp}).
When building the union graph (Phase~3), for $q=1,\ldots,n$, the $\LCS$ between two consecutive nodes, say $v_i$ and $w_j$, is equal to the $\LCP$ of their reverses $\reprr{v_i}$ and $\reprr{w_j}$, which is given by $B[q]-1$ whenever $B[q]>0$ (if $B[q]=0$ then $\reprr{v_i}=\reprr{w_j}$ and nodes $v_i$ and $v_j$ should be merged).  Hence, our algorithm for computing the VO representation of the union graph consists exactly of the algorithm in Fig.~\ref{fig:xHMalgo} in which we store the array $B$ in $n\log k$ bits instead of using the 2-bit representation described in Section~\ref{s:implementation}. Hence the running time is still $\Oh(m k)$ and the  working space becomes the space for the bitvectors $\bv{h-1}$ and $\bv{h}$ (recall we define the working space as the space used in addition to the space for the input and the output). 

\begin{theorem}
Merging two succinct representations of variable order \dbG\ graphs takes $\Oh(\edges k)$ time and $2n$ bits plus $\Oh(\sigma)$ words of working space.\qed
\end{theorem}

Note the $\LCP$ values can be written directly to disk, using for example the technique from~\cite{almob/EgidiLMT19}.

\section{Merging \WGs via 2-SAT}\label{sec:2sat}

Merging two \dbG graphs $G_0$ and $G_1$, or other succinct indices~\cite{tcs/EgidiM20}, amounts to building a new succinct data structure that supports the retrieval of the elements which are in $G_0$ or in $G_1$. Because of the correspondence between succinct data structures and Wheeler graphs, the natural generalization of the problem of merging succinct indices is the following problem: 

\begin{problem}\label{prob:1}
Given two \WAs\ $\Auz = (V_0,E_0,F_0,s_0,\Sigma)$ and $\Auo = (V_1,E_1,F_1,s_1,\Sigma)$ recognizing respectively the languages \langz and \lango find a \WA $\Aux$ recognizing the union language $\langx = \langz \cup \lango$, or report that none exists.
\end{problem}

Unfortunately, in the general case we cannot even guarantee that a \WA recognizing $\langx$ exists, \replaced{since the property of languages of being recognizable by a Wheeler automata is not closed under union~\cite[Lemma 3.3]{AlankoCorr2020}}{
(see~\cite[Theorem~2.1]{Alanko2020} for some results in this area)}. Hence, instead of tackling the general problem, in this paper \added{we reason in terms of automata rather than languages}.  Starting from $\Auz$ and $\Auo$ we define the union automaton $\ua$ that naturally recognizes $\langx$ and we consider the problem of determining if there exists an ordering of $\ua$'s nodes that makes it a \WG (with an additional requirement detailed below). 
Formally, we define the union automaton (or graph) $\ua = (V,E,F,s,\Sigma)$ as follows:
\begin{enumerate}
    \item  $V = (V_0\setminus\{s_0\}) \cup (V_1\setminus\{s_1\}) \cup \{s\}$;
    \item $E=E_0^* \cup E_1^*$ where, for $i=0,1$, $E_i^*$ is $E_i$ where each edge leaving $s_i$ is replaced by an edge leaving $s$ with the same destination;
    \item $F = (F_0\setminus\{s_0\}) \cup (F_1\setminus\{s_1\})$; with $s$ added to $F$ if $s_0\in F_0$ or $s_1 \in F_1$. 
\end{enumerate}

It is immediate to see that $\ua$ recognizes $\langx = \langz \cup \lango$ and still has the property that the initial state $s$ is the only one with in-degree 0. 

\begin{definition} \label{def:wo}
A {\em Wheeler \added{Compatible} Order} (\WO) for the union automaton $\ua$ is an ordering of $V$ that makes $\ua$ a Wheeler graph and that is compatible with the orderings of $V_0$ and $V_1$, in the sense that if $u,v \in V_0\setminus\{s_0\}$ (resp. $V_1\setminus\{s_1\}$) then $u<v$ in $V$ iff $u<v$ in $V_0$ (resp. $V_1$).\qed
\end{definition}

\begin{figure}[t]
    \centering
    \includegraphics[width=0.7\textwidth]{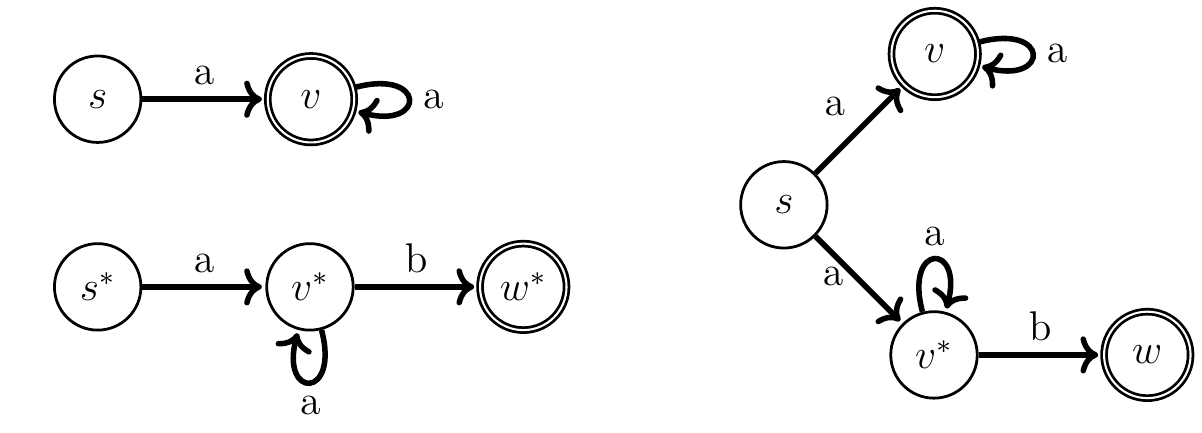}
    \caption{Two Wheeler automata (left) and their union (right). The accepting states are denoted, as usual, with a double circle. The union automaton is not Wheeler (see text), but if we make $v^*$ final the automaton on the lower left accepts the union language that is therefore Wheeler.}\label{f:automata}
\end{figure}

In the rest of this section, we consider the problem of determining whether $\ua$ has a \WO and, if this is the case, to explicitly find it:

\begin{problem}\label{prob:2}
\added{Given the union automaton \ua of two \WAs, as defined above, find a \WO of \ua in the sense of Definition~\ref{def:wo} or report that none exists.}
\end{problem}

\added{We observe that a \WO does not necessarily exist even if the union language is Wheeler. Fig.~\ref{f:automata} shows two Wheeler automata (left) and their union automata (right). The two input automata recognize respectively the languages $a^n$ and $a^n b$ with $n\geq 1$. The automaton on the right recognizes the union language but it is not a Wheeler automaton: clearly by~\eqref{eq:wgA} it must be $s < w$ and the ranks of $v$ and $v^*$ must be between $s$ and $w$ but we cannot find {ranks for $v$ and $v^*$ in the ordering that satisfy} the Wheeler conditions. Indeed, $s < v^*$ and the edges $(s,v,a)$ and $(v^*,v^*,a)$ implies $v < v^*$; similarly $s < v$ implies $v^* < v$. However, the union language is Wheeler, since it is recognized by the second input automaton (left) with the state $v^*$ made final.}

\added{Having established its limitations, in the rest of this section we provide a solution to Problem~\ref{prob:2}.} It is known~\cite{Gibney2019} that in the general case determining if an automaton is a \WA is NP-complete. However, for the union automaton \ua we show there exists a polynomial time algorithm to determine whether it has a \WO. The algorithm provides a \WO if one exists and works by transforming our problem into a 2-SAT instance that can then be solved using any polynomial time 2-SAT solver ({\it e.g.} \cite{AspvallPT79}). The idea of using 2-SAT for recognizing \WAs was introduced in~\cite{Alanko2020}; however, in the general case this approach is viable only if each node has at most two outgoing edges labeled with the same symbol~\cite[Theorem~3.1]{Alanko2020}. We show that this limitation can be removed in the special case that the input is the union automaton of two \WAs.

Let $s, u_1, u_2, \ldots, u_n, v_1, v_2, \ldots v_m$ denote the nodes of the union automata, where {$u_1 < u_2 < \cdots < u_n$} are the ordered nodes of $\Auz$ and   $v_1 < v_2 < \cdots < v_m$ are the ordered nodes of $\Auo$. We need to check if it is possible to ``merge'' the two sets of ordered nodes so that the Wheeler conditions are met. Our strategy consists in building a set of clauses, with at most 2 literals each, and to show that there exists a \WO if and only if all clauses can be satisfied simultaneously. 

We introduce $nm$ boolean variables $\lij{i}{j}$  for $i=1,..,n$, $j=1,..,m$, where $\lij{i}{j}$ represents the condition $(u_i < v_j)$. We introduce a first set of clauses:
\begin{align}
    i=1,..,n,\; j=1,..,m-1 \qquad \lij{i}{j} &\Rightarrow \lij{i}{j+1}\label{eq:complete} \\
    i=2,..,n,\; j=1,..,m \qquad \lij{i}{j} &\Rightarrow \lij{i-1}{j} \\
    i=1,..,n,\; j=2,..,m \quad\gij{i}{j} &\Rightarrow \gij{i}{j-1}\label{eq:complete_xlast} \\
   i=1,..,n-1,\; j=1,..,m  \quad\gij{i}{j} &\Rightarrow \gij{i+1}{j}\label{eq:complete_last}
\end{align}
Informally, these clauses ensure the transitivity of the resulting order (note that \eqref{eq:complete_xlast} for example is equivalent to $(u_i > v_j) \Rightarrow (u_i > v_{j-1})$). Next, we introduce a second set of clauses which ensure that the resulting order makes $\ua$ a \WA. For each pair of nodes $u_i \in V_0$, $v_j \in V_1$ such that the edges entering $u_i$ are labeled $a$ and the edges entering $v_j$ are labeled $a'$ with $a\neq a'$ we add the clause
\begin{equation}\label{eq:wgb:clause}
\begin{cases}
\lij{i}{j} &\mbox{if } a<a' \\
\gij{i}{j} &\mbox{if } a>a'
\end{cases}
\end{equation}
which is equivalent to~\eqref{eq:wgA}. Finally, for each symbol $a$ and for every pair of edges $(u_i,u_{k}) \in E_0$ and $(v_j,v_{h}) \in E_1$ both labeled $a$ we add the clauses
\begin{align}
&\lij{i}{j} \Rightarrow \lij{k}{h}\\
&\gij{i}{j} \Rightarrow \gij{k}{h}\label{eq:last_clause}.
\end{align}
which are equivalent to~\eqref{eq:wgB} (note we cannot have $u_k=v_h$). 

\begin{lemma}
A truth assignment for the variables $\lij{i}{j}$ that satisfies~\eqref{eq:complete}--\eqref{eq:last_clause} induces a \WO for the nodes of~$\ua$. Viceversa, a \WO for \ua provides a solution for the 2-SAT instance defined by clauses~\eqref{eq:complete}--\eqref{eq:last_clause}.  
\end{lemma}

\begin{proof}
Given an assignment satisfying~\eqref{eq:complete}--\eqref{eq:last_clause} consider the ordering of $V$ defined as follows. Node $s$ has the smallest rank, the nodes in $V \cap V_0$ (resp. $V \cap V_1$) have the same order as in $V_0$ (resp. $V_1)$ and for each pair $u_i \in V_0$, $v_j \in V_1$ it is $u_i < v_j$ iff $\lij{i}{j}$ is true. The resulting order is total: the only non trivial condition being the transitivity which is ensured by clauses~\eqref{eq:complete}--\eqref{eq:complete_last}. In addition, the order makes \ua a \WA since conditions~\eqref{eq:wgA}--\eqref{eq:wgB} follow by the hypothesis on $\Auz$ and $\Auo$ if the edges $e$ and $e'$ both belong to $E_0$ or $E_1$, and by~\eqref{eq:wgb:clause}--\eqref{eq:last_clause} if not.
Viceversa, given a \WO for $V$ it is straightforward to verify that the assignment $\lij{i}{j} = (u_i < v_j)$ for $i=1,\ldots,n$, $j=1,\ldots, m$ satisfies the clauses~\eqref{eq:complete}--\eqref{eq:last_clause}.\qed 
\end{proof}

The following theorem summarizes the results  of this section. 

\begin{theorem} \label{theo:2sat}
Given two \WAs $\Auz$ and $\Auo$ in $\Oh(|E_0| |E_1|)$ time we can find a \WO for the union automata or report that no such order exists.
\end{theorem}

\begin{proof}
The construction of the clauses takes constant time for clause and there are $\Oh(|V_0||V_1|)$ clauses of type~\eqref{eq:complete}--\eqref{eq:complete_last}, and $\Oh(|E_0| | E_1|)$ clauses of type~\eqref{eq:wgb:clause}--\eqref{eq:last_clause}. The thesis follows observing that a 2-SAT instance can be solved in linear time in the number of clauses~\cite{AspvallPT79}.\qed
\end{proof}

\section{Computing a \WO by iterative refining}\label{sec:refining}

The major drawback of the algorithm of Section~\ref{sec:2sat} is its large working space. The explicit construction of 2-SAT clauses will take much more space than the succinct representation of the input/output automata.
As discussed in Section~\ref{sec:stateoftheart} for \dbG graphs, space has a significant practical impact; hence a possible line of future research could be to maintain an implicit representation of the clauses and to devise a memory efficient 2-SAT solver. 

In this section we present a different algorithm for computing a \WO that takes $\Oh(|V|^2)$ time and only uses $4|V| + o(|V|)$ bits of working space. Our algorithm however does not always compute a \WO for the union automata \ua. Instead, under the assumption that a \WO for \ua exists, our algorithm returns a \WA \uax, possibly different from \ua, that recognizes the same language as \ua. The automaton \uax is always smaller than or equal to \ua and the algorithm explicitly returns also the ordering that makes \uax a \WA. Interestingly it is even possible that our algorithm returns a \WA recognizing the union language $\langz \cup \lango$ even if no \WO for \ua exists. \added{This is a positive feature, but implies that our algorithm is not solving Problem~\ref{prob:2}, but rather it is providing an (unfortunately) incomplete solution to Problem~\ref{prob:1}. We will discuss this point in detail at the end of 
Section~\ref{ss:iteref}.}

To describe our algorithm we introduce some additional notation. 

\begin{definition}\label{def:wcons}
Let $V$ denote the set of states of the union automata $\ua$. {An ordered} partition $\Bx_0, \Bx_1, \ldots \Bx_k$ of~$V$ into disjoint subsets is said to be {\em \Wcons} if $x\in\Bx_i$, $y\in\Bx_j$ with $i<j$ implies that for any \WO it is $x<y$.\qed
\end{definition}

In the above definition it is clear that the ordering of the sets in the partition is fundamental; however for simplicity in the following we usually leave implicit that we are talking about ordered partitions. Because of condition~\eqref{eq:wgA}, in a \WG all edges entering a given node $v$ must have the same label; this observation justifies the following definition.

\begin{definition}\label{def:lambda}
If $v$ is a node in a \WG with positive in-degree, we denote by $\lab{v}$ the symbol labelling every edge entering in~$v$.\qed
\end{definition}

\noindent
With the above notation, the simplest example of a \Wcons partition is given by the following lemma.

\begin{lemma}\label{lemma:startWcons}
Let $\Bx_{0} = \{s\}$, and for $i=1,\ldots,\sigma$, let $\Bx_i = \{ v\in V | \lab{v}=i\}$. Then, $\Bx_{0},\Bx_1, \ldots,\Bx_{\sigma}$ is a  \Wcons partition.
\end{lemma}

\begin{proof}
$\Bx_{0}, \Bx_1, \ldots,\Bx_{\sigma}$ is a well defined partition since $s$ is the only state with in-degree 0 and $\Auz$, $\Auo$ are \WA. The partition is \Wcons because any \WO for \ua must satisfy property~\eqref{eq:wgA}.\qed 
\end{proof}

The following lemmas illustrate some useful properties of \Wcons partitions. 

\begin{lemma}\label{lemma:WGm0}
Let $\Bx_0, \Bx_1, \ldots \Bx_k$ denote a \Wcons partition and let $v,v'\in \Bx_h$ be such that $v\neq v'$ and $\lab{v}=\lab{v'}$. If there exist two edges $(u,v)$ and $(u',v')$ with $u\in\Bx_i$, $u'\in \Bx_j$, $i<j$, then in any \WO we must have $v<v'$.
\end{lemma}

\begin{proof}
By Definition~\ref{def:wcons} in any \WO we must have $u< u'$; the thesis follows by~\eqref{eq:wgB}.\qed
\end{proof}

\begin{lemma}\label{lemma:WGm2}
Let $\Bx_0, \Bx_1, \ldots \Bx_k$ denote a \Wcons partition and let $v,v'\in \Bx_h$ be such that $v\neq v'$ and $\lab{v}=\lab{v'}$. Let $\ell$ (resp. $\ell'$) denote the smallest index such that there exists an edge from a node in $\Bx_\ell$ (resp. $\Bx_{\ell'}$) to $v$ (resp. $v'$). Similarly,  let $m$ (resp. $m'$) denote the largest index such that there exists an edge from a node in $\Bx_m$ (resp. $\Bx_{m'}$) to $v$ (resp. $v'$). If it is not $\ell=m=\ell'=m'$, then, for any \WO we have:
\begin{subequations}
\begin{align}
    m \leq \ell' & \quad\Longrightarrow\quad v < v' \label{eq:mell0}\\
    m' \leq \ell & \quad\Longrightarrow\quad v > v' \label{eq:mell1}.
\end{align}
\end{subequations}
In addition, if it is not $(m \leq \ell') \vee (m' \leq \ell)$, then a \WO for the union automaton cannot exist.
\end{lemma}

\begin{proof}
Consider the case $m \leq \ell'$ ($m' \leq \ell$ is symmetrical). If $m<\ell'$ then $v<v'$ by Lemma~\ref{lemma:WGm0}. If $m=\ell'$, since {we are assuming} it is not $\ell=m=\ell'=m'$ we must have either $\ell<\ell'$ or $m<m'$ (or both). In all cases the thesis follows again by Lemma~\ref{lemma:WGm0}.

Suppose now that it is not $(m \leq \ell') \vee (m' \leq \ell)$; then we must have $(m > \ell') \wedge (m' > \ell)$. Again by Lemma~\ref{lemma:WGm0} a \WO should satisfy simultaneously $v>v'$ and $v'>v$ which is impossible.\qed
\end{proof}

In the following we call the $(\ell,m)$ pair defined in Lemma~\ref{lemma:WGm2} a {\em \minmax} pair.

\begin{definition}\label{def:compat}
We say that two \minmax pairs $(\ell,m)$ and $(\ell',m')$  are {\em compatible} if $(m \leq \ell') \vee (m' \leq \ell)$.\qed
\end{definition}

With the above definition, we can rephrase the second half of Lemma~\ref{lemma:WGm2} saying that if the \minmax pairs $(\ell,m)$ and $(\ell',m')$ are not compatible then the union automaton does not have a \WO. 

Given two {\em compatible} \minmax pairs $(\ell,m)$ and $(\ell',m')$ if $m\leq \ell'$ we write
\begin{equation}\label{eq:mmorder}
(\ell,m) \preceq (\ell',m').
\end{equation}
It is easy to see that if $(\ell,m)$ and $(\ell',m')$ are compatible then it is either $(\ell,m) \preceq (\ell',m')$ or $(\ell',m') \preceq (\ell,m)$ and both relations are true simultaneously if and only if $\ell=m=\ell'=m'$. Also note that the relation $\preceq$ is transitive in the sense that if $(\ell,m) \preceq (\ell',m')$ and $(\ell',m') \preceq (\ell'',m'')$ then $(\ell,m)$ is compatible with $(\ell'',m'')$ and $(\ell,m) \preceq (\ell'',m'')$. 

\begin{figure}[t]
\hrule\smallbreak
\begin{algorithmic}[1]
\State $\newP \gets \{ \Bx_0 \}$  \Comment{Init new partition with $\Bx_0$}
\For{$i \gets 1$ \KwTo $k$} \Comment{Consider $\Bx_1,\ldots,\Bx_k$}
  \State $\Bz \gets \Bx_i \cap V_0$ \Comment{Nodes coming from $\Auz$}\label{line:loop0}
  \State $\Bo \gets \Bx_i \cap V_1$ \Comment{Nodes coming from $\Auo$}
  \State $L \gets \mergex(\Bz,\Bo)$ \Comment{Merge according to \minmax pairs ordering}\label{line:merge}
  \State Split $L$ into subsets $L_1, \ldots, L_t$   with identical \minmax pairs.\label{line:split}
  \State $\newP \gets \newP \cup \{L_1, \ldots, L_t\}$ \Comment{Add $\Bx_i$'s refinement to \newP}\label{line:loop1}
\EndFor
\State\textbf{return} \newP
\end{algorithmic}
\smallbreak\hrule
\caption{Refinement step for a \Wcons\ partition  $\Bx_0,\Bx_1,\ldots,\Bx_k$. It returns a refined partition \newP unless during the merging step two incompatible \minmax pairs are found; in this case the algorithm terminates reporting that no \WO exists for~$\ua$.}
\label{fig:refine}
\end{figure}

\subsection{The iterative refining algorithm}\label{ss:iteref}

Our strategy consists in starting with the \Wcons partition defined in Lemma~\ref{lemma:startWcons} and refining it iteratively obtaining finer \Wcons partitions. Refining a partition $\Bx_0, \Bx_1, \ldots, \Bx_k$ here means that each set $\Bx_i$ is partitioned into $\Bx_{i,0}, \Bx_{i,1}, \ldots, \Bx_{i,t_i}$ so that at the next iteration the working partition becomes
$$
\Bx_{0,0}, \ldots, \Bx_{0,t_0},
\Bx_{1,0}, \ldots, \Bx_{1,t_1} \,\ldots\,
\Bx_{i,0}, \ldots, \Bx_{i,t_i} \,\ldots\,
\Bx_{k,0}, \ldots, \Bx_{k,t_k}.
$$
To simplify the notation we use $\Bx_0, \Bx_1, \ldots \Bx_k$ to denote the current partition assuming that after each refinement step the sets are properly renumbered.

At each iteration the refinement is done with the algorithm outlined in Fig.~\ref{fig:refine} that produces a finer partition or reports that no \WO exists. We first observe that since we start with the partition given by Lemma~\ref{lemma:startWcons}, at any time during the algorithm each set $\Bx_h$ is a subset of some 
$\{ v\in V | \lab{v}=i\}$,  
so we can apply Lemmas~\ref{lemma:WGm0} and~\ref{lemma:WGm2} to any pair of distinct nodes $v,v'\in\Bx_h$.  

In the main loop of the algorithm in Fig.~\ref{fig:refine}, given a set $\Bx_i$ we define $\Bz = \Bx_i\cap V_0$ and $\Bo = \Bx_i \cap V_1$. Since $\Auz$ (resp. $\Auo$) is a \WG the elements in $\Bz$  (resp. $\Bo$) are all pairwise compatible and they are already ordered according to relation $\preceq$ defined by~\eqref{eq:mmorder}. Next we merge $\Bz$ and $\Bo$
according to~ $\preceq$. We start with an empty result list $L$ and we compare the nodes currently at the top of $\Bz$ and $\Bo$, say $v\in\Bz$ and $v'\in\Bo$. If the corresponding \minmax pairs $(\ell,m)$ and $(\ell',m')$ are not compatible the whole algorithm fails (no \WO exists by Lemma~\ref{lemma:WGm2}). Assuming the pairs are compatible, if $(\ell,m) \preceq (\ell',m')$ we remove $v$ from $\Bz$ and append it to $L$; otherwise we remove $v'$ from the $\Bo$ and append it to $L$. 
In either case, we then continue the merging of the elements still in $\Bz$ and $\Bo$. At the end of the merging procedure all elements in $\Bx_i$ are in $L$ ordered according to the $\preceq$ relation. We refine $\Bx_i$ splitting $L$ into (maximal) subsets $L_1, L_2, \ldots, L_t$ so that all elements in the same subset $L_k$ have identical \minmax pairs $(\ell,m)$. In other words, if $v\in L_j$ and $v' \in L_k$, with $j<k$, the corresponding pairs $(\ell,m)$ and $(\ell',m')$ are such that $(\ell,m) \preceq (\ell',m')$ but it is not $\ell=m=\ell'=m'$. By~\eqref{eq:mell0} this implies that in any \WO we must have $v < v'$ and this ensures that if we split $\Bx_i$ into the subsets $L_1, L_2, \ldots, L_t$ the resulting partition is still \Wcons. Repeating the above procedure for every set $\Bx_i$ we end up with either a refined partition or the indication that the union automaton $\ua$ has no \WO.  In the former case it is also possible that the new partition is identical to the current one: this happens when for $i=1,\ldots,k$ at line~\ref{line:split} it is $t=1$ since all nodes in $L$ have the same \minmax pairs. In this case no further iteration will modify the partition so we stop the refinement phase.

\begin{figure}[h]
\centering
\begin{minipage}[b]{.5\textwidth}
  \begin{flushleft}
	\includegraphics[width=\columnwidth]{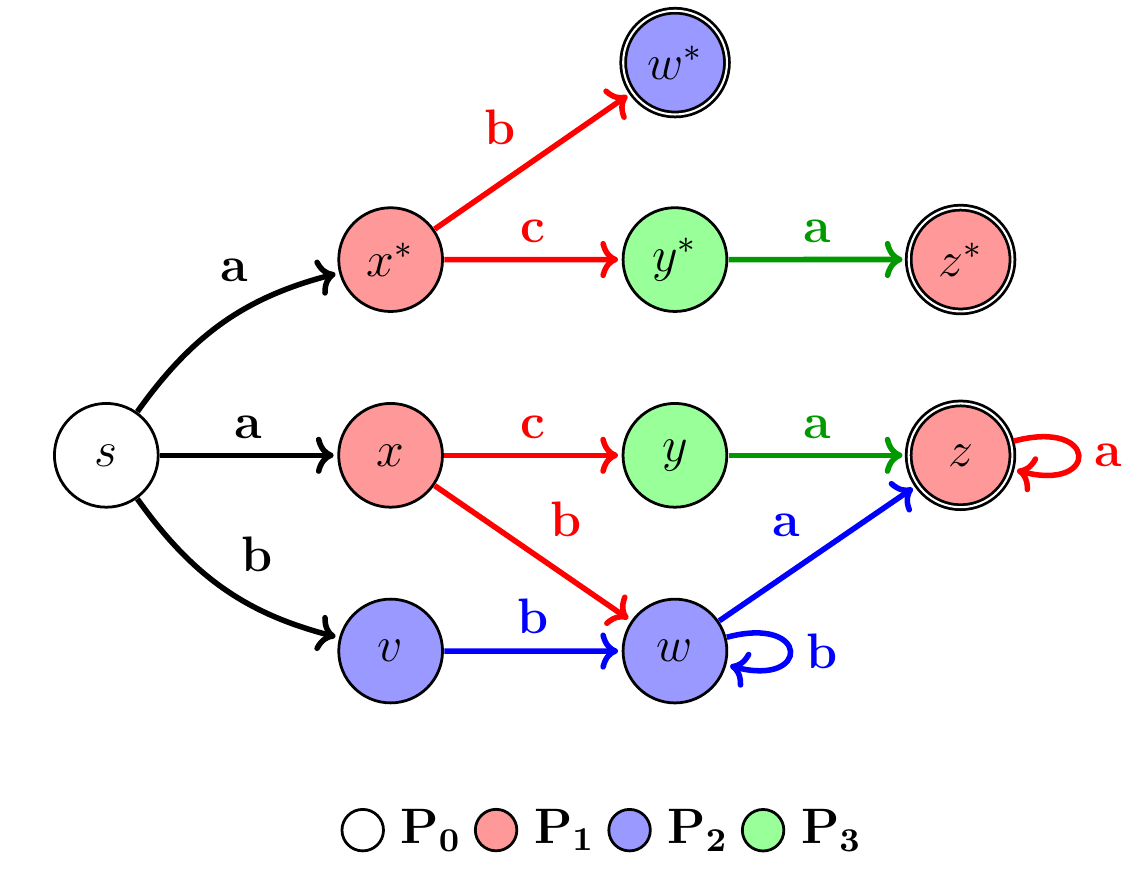}
 \end{flushleft}
 \vfill
 \end{minipage}%
\begin{minipage}[t]{.5\textwidth}
  \begin{flushright}
\includegraphics[width=\columnwidth]{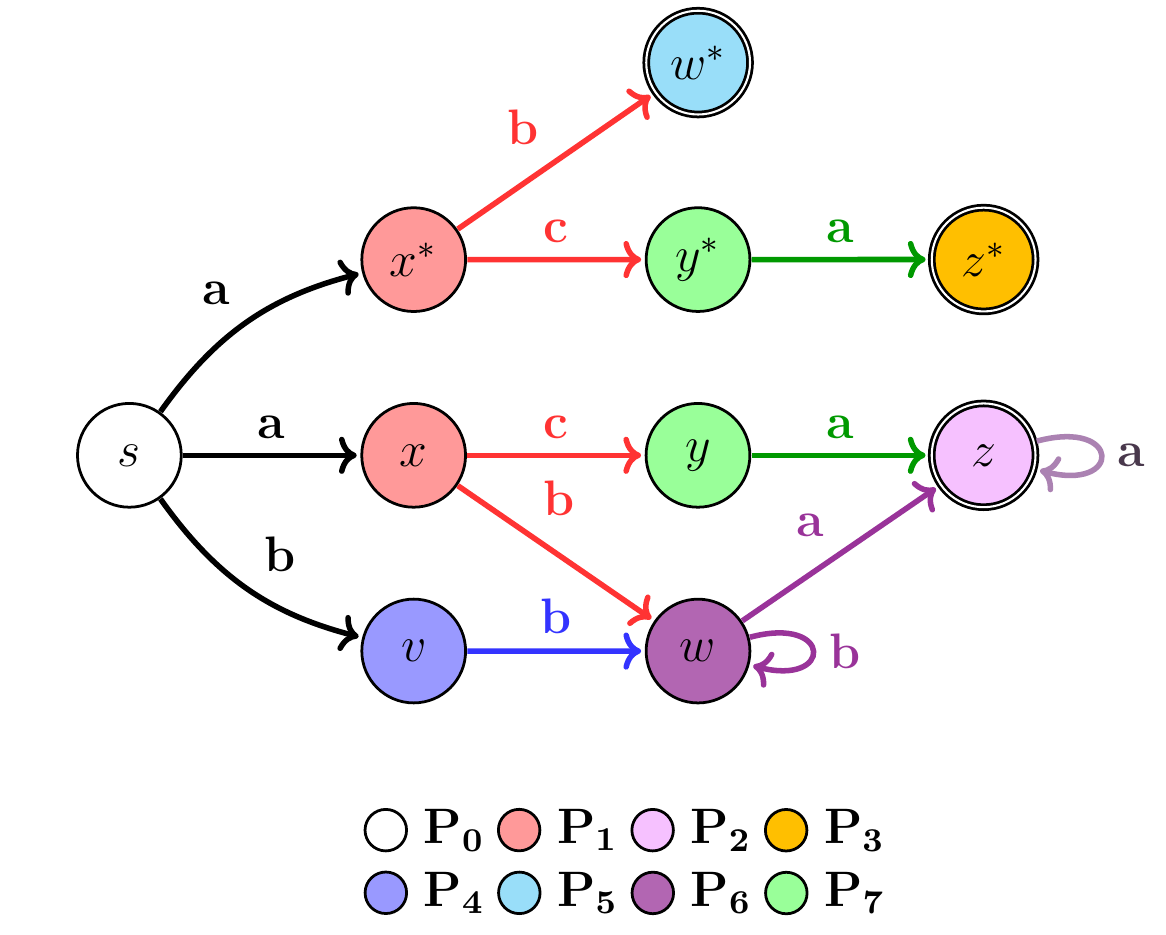}
 \end{flushright}
  \end{minipage}%
  \caption{Example of a refinement step for the union automaton displayed above obtained from the \WAs 
    $\Auz$ and $\Auo$, with $V_0 = \{s, x^*, y^*, w^*, z^*\}$ and $V_1 =  \{s, x, y, v, w, z\}$
    On the left, nodes are colored to highlight the initial partition from Lemma~\ref{lemma:startWcons}: $\Bx_0 = \myset{s}$, $\Bx_1 = \myset{x^*,z^*,x,z}$, $\Bx_2=\myset{w^*, v, w}$, $\Bx_3=\myset{y^*,y}$. There are edges going from $\Bx_0$ to $\Bx_1$ and $\Bx_2$, from $\Bx_1$ to $\Bx_2$ and $\Bx_3$, from $\Bx_2$ to $\Bx_1$ and $\Bx_2$, and from $\Bx_3$ to $\Bx_1$ (colors in the figure highlight the partition set from which edges originate). As a result, at the end of the refinement step, the new partition is $\Bx_0 = \myset{s}$, $\Bx_1 = \myset{x^*,x}$, $\Bx_2 = \myset{z}$, $\Bx_3 = \myset{z^*}$, $\Bx_4=\myset{v}$, $\Bx_5=\myset{w^*}$ $\Bx_6=\myset{w}$ $\Bx_7=\myset{y^*,y}$, as shown on the right side. An additional refinement step does not further modify the partition since all sets are either singletons or all their nodes have the same \minmax pairs.}
    \label{f:pathological}
\end{figure}

Fig.~\ref{f:pathological} shows an example of a refinement step. $\Bx_1$ on the left is refined yielding $\Bx_1-\Bx_3$ on the right, and $\Bx_2$ on the left is refined yielding $\Bx_4-\Bx_6$ on the right. All other $\Bx_i$'s are unchanged. The partition on the right cannot be further refined.

\mysubsubsection{Bounding the number of iterations}
Recall we are assuming that each node in the union automata is reachable from the initial state $s$. For each node $v\in V$ we define $d_v$ as the length of the {\em shortest path} from $s$ to $v$. Let then
\begin{equation}\label{eq:deltadef}
    \delta = \max_{v\in V} d_v   
\end{equation}
be the maximum distance from $s$ to nodes in~$V$
In the following we show that, unless our algorithm reports that there is no \WO, after at most $\delta+2$ refinement iterations we reach a \Wcons partition that is not further refined by the algorithm in Fig.~\ref{fig:refine}. 

\begin{lemma}\label{lemma:split}
If, at the beginning of the $j$-th iteration, a partition element $\Bx_i$ is not a singleton then either: $(i)$ for every $v\in\Bx_i$ all paths from $s$ to $v$ have length at least $j$ or $(ii)$ there exists $j'<j$ such that for every $v\in\Bx_i$ all paths from $s$ to $v$ have length $j'$. 
\end{lemma}

\begin{proof}
We prove the result by induction on $j$. For $j=1$, immediately before the first iteration for $i>1$ each $\Bx_i$ is defined as $\Bx_i = \{v | \lab{v}=i\}$
so they all satisfy property $(i)$. 
For $j>1$, let $\Bx_0, \ldots, \Bx_k$ denote the current partition at the beginning of the $(j-1)$-st iteration. During the refinement step each set $\Bx_i$ is split into subsets $L_1,L_2,\ldots,L_t$ as described above. Each subset $L_h$ will become a partition element for the $j$-th iteration, so to prove the lemma we need to show that $L_h$ satisfies $(i)$ or $(ii)$. By construction, if $L_h$ is not a singleton then all nodes $v\in L_h$ have the same \minmax pair $(\ell,m)$ with $\ell=m$. It follows that {\em all} edges reaching the nodes in $L_h$ must originate by the same set $\Bx_{\ell}$. If $\ell=0$ then $L_h$ satisfies property $(ii)$ with $j'=1$. If $\ell>0$, by induction {$\Bx_{\ell}$} must satisfy either $(i)$ or $(ii)$. If $\Bx_{\ell}$ satisfies $(i)$ and all paths from $s$ to $\Bx_{\ell}$ have length at least $j-1$, then $L_h$ also satisfies $(i)$ with paths of length at least $j$; if $\Bx_{\ell}$ satisfies $(ii)$ with length $j'$, then $L_h$ satisfies $(ii)$ with length $j'+1$.\qed
\end{proof}

\begin{lemma}\label{lemma:nosplit}
If, at the beginning of the $j$-th iteration, a partition element $\Bx_i$ is not a singleton and satisfies property $(ii)$ of Lemma~\ref{lemma:split}, then it will not be split by all subsequent refinement steps.
\end{lemma}

\begin{proof}
We prove the result by induction on $j$. For $j=1$ there cannot be any $\Bx_i$ satisfying property $(ii)$ of Lemma~\ref{lemma:split}. To prove the lemma for $j=2$ we observe that, by the proof of Lemma~\ref{lemma:split}, during the first iteration a set satisfying property $(ii)$ is generated only by a subset $L_h$ containing only nodes with \minmax pair $(0,0)$ (that is, nodes only reachable from the source in one step). In any further refinement step the \minmax pair of each node will still be $(0,0)$ so the set will not be further modified. 

For $j>2$ we observe again that, by the proof of Lemma~\ref{lemma:split}, in any subsequent iteration a new partition element satisfying property $(ii)$ is generated only when all nodes in a subset $L_h$ have the same \minmax pair $(\ell,\ell)$ and $\Bx_\ell$ is a set already satisfying property $(ii)$. By inductive hypothesis the set $\Bx_\ell$ will not be further split in subsequent iterations; hence the nodes in $L_h$ will still have identical \minmax pairs $(\ell',\ell')$ in subsequent iterations and the subset $L_h$ will not be further split.\qed
\end{proof}

We use Lemmas~\ref{lemma:split} and~\ref{lemma:nosplit} to bound the number of refinement steps in terms of the maximum distance~\eqref{eq:deltadef}:

\begin{lemma} \label{lemma:delta}
Let $\delta$ be as defined in~\eqref{eq:deltadef}. After at most $\delta+2$ refinement iterations either the algorithm has reported that a \WO does not exist, or it has computed a \Wcons partition that the algorithm could not refine.
\end{lemma}

\begin{proof}
After $\delta+1$ refinement iterations there cannot be a  partition element $\Bx_i$ satisfying property $(i)$ of Lemma~\ref{lemma:split} since $\delta$ is the maximum distance of any node from $s$.
Hence, after at most $\delta+1$ iterations all partition elements $\Bx_i$ are either singletons or satisfy property $(ii)$ of Lemma~\ref{lemma:split}; by Lemma~\ref{lemma:nosplit} none of them will be refined in subsequent iterations.\qed
\end{proof}

\mysubsubsection{Construction of the Wheeler Automaton for the union language} When the partition $\Bx_0,\Bx_1,\ldots,\Bx_k$ cannot be further refined we proceed building the output automaton. One can see that if all sets $\Bx_i$ are singleton, then the partition ordering is a \WO for \ua. In the general, case in which some $\Bx_i$ is not a singleton, we use the partition to build a (smaller) \WA that also recognizes the union language.

\begin{definition}\label{def:finalaut}
Let $\ua = (V,E,F,s,\Sigma)$ be the union automaton and $\{\Bx_0,\Bx_1,\ldots,\Bx_k\}$ a \Wcons partition of its nodes that cannot be further refined. We define the automaton $\uax = (V',E',F',s,\Sigma)$ with $V' = \{\Bx_0,\Bx_1,\ldots,\Bx_k\}$, $s = \Bx_0$,  $(\Bx_i,\Bx_j, a)\in E'$ iff $(v,v', a)\in E$ for some $v\in \Bx_i$ and $v'\in \Bx_j$, and $\Bx_i\in F'$ iff $\Bx_i\cap F\neq \emptyset$. \qed
\end{definition}

The automaton is well defined, since edges incoming in each $\Bx_i$ have the same label in \ua and therefore there is no ambiguity in the definition of edge labels in $\uax$. 

\begin{lemma} \label{lemma:samelang}
$\uax$ recognizes the same language as $\ua$.
\end{lemma}
\begin{proof}
To prove the lemma we preliminary observe that if $v, v'\in\Bx_i$ are both nodes from the same input automaton, say $\Auz$, then they are equivalent in the sense that if there is in $\Auz$ a path with label $\alpha$ from $s$ to $v$, there is in $\Auz$ also a path with the same label from $s$ to $v'$, and vice versa.
Therefore we can safely assume that each $\Bx_i$  is either singleton or contains a node from $V_0$ and a node from $V_1$; for simplicity here we call these nodes {\em special nodes}.

Since the partition $\{\Bx_0,\Bx_1,\ldots,\Bx_k\}$ cannot be further refined we get that in \ua all edges entering into a $\Bx_h$ which is not a singleton originate from a single set $\Bx_{h'}$, which therefore either contains nodes from both $V_0$ and $V_1$ or contains the node $s$ (when $h'=0$). In \uax this implies that each special node is reachable by a single node which is either $s$ or a special node itself. Hence, \uax's subgraph containing the special nodes and the source $s$ is a tree rooted at $s$. 

Consider now the word $\alpha$ corresponding to a path in \uax going from $s$ to a final node $f \in F'$. If $f$ is a special node all nodes in the path are special. By construction, the partition element $\Bx_f$ in $\ua$  corresponding to  $f$ in $\uax$ must contain a node from either $F_0$ or $F_1$ (or both), hence a path labeled $\alpha$ is contained in either  $\Auz$ or $\Auo$ and is therefore contained in the union language. If $f$ is not a special node, then a proper prefix of the path from $s$ to $f$ consists of special nodes, while $\Bx_f\cap F_0\neq\emptyset$ or  $\Bx_f\cap F_1\neq\emptyset$, but not both. Assuming for example $\Bx_f\cap F_0\neq\emptyset$, then we can find a path in  $\Auz$
labeled $\alpha$ which will therefore belong to the union language. 

Finally, if the word $\beta$ corresponds to a path in \ua from $s$ to a final node, by possibly mapping some nodes in \ua to the corresponding special nodes in \uax we get a path labeled $\beta$ in \uax.\qed
\end{proof}

\begin{theorem}
\uax is a \WA recognizing the union language $\langz \cup \lango$.
\end{theorem}

\begin{proof}
By Lemma~\ref{lemma:samelang}, we only need to prove that there is an ordering of the nodes of $\uax$ that satisfies the Wheeler conditions. 

Consider the ordering $\Bx_0<\Bx_1<\cdots<\Bx_k$ and assume by contradiction that this ordering does not satisfy the Wheeler conditions. Condition~\eqref{eq:wgA} is verified by construction by the initial partition (the one from Lemma~\ref{lemma:startWcons}), and all subsequent refinements never change the relative order of existing partition elements, since they are just split in smaller subsets. If condition~\eqref{eq:wgB} is not verified the graph has edges $(\Bx_i,\Bx_h,a)$ and  $(\Bx_j,\Bx_\ell,a)$ such that $\Bx_j> \Bx_i$ and $\Bx_h > \Bx_\ell$. However, we notice that at the beginning of the refinement algorithm the nodes in $\Bx_h$ and $\Bx_\ell$ were in the same partition element $\{v | \lab{v}=a \}$. At the iteration $\tau$ in which these nodes ended in different sets, the nodes in $\Bx_\ell$ had a \minmax pair smaller, according to~\eqref{eq:mmorder}, than the nodes in $\Bx_h$. This implies that no edge reaching $\Bx_\ell$ originated from a partition element {\em following} the partition elements originating the edges reaching $\Bx_h$. This was true at iteration $\tau$ but since the refinement process splits partition elements but never changes their relative order it is impossible that $\Bx_j > \Bx_i$.\qed
\end{proof}

\begin{figure}[h]
\centering
\begin{minipage}[b]{.5\textwidth}
  \begin{flushleft}
\includegraphics[width=\columnwidth]{sndPart.pdf}
 \end{flushleft}
 \end{minipage}%
\begin{minipage}[b]{.5\textwidth}
  \begin{flushright}
\includegraphics[width=\columnwidth]{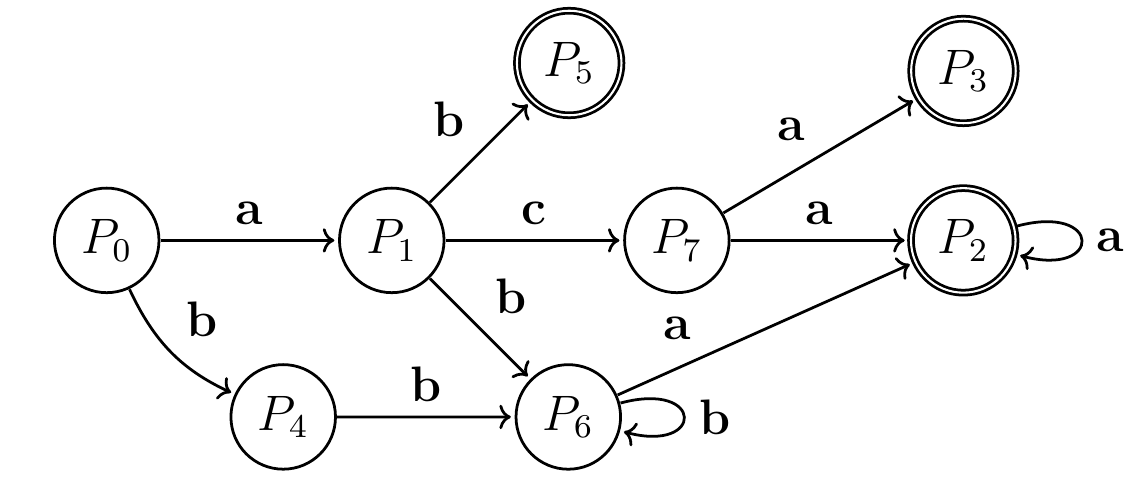}
 \end{flushright}
  \vspace{9mm}
  \end{minipage}%
  \caption{The Wheeler automaton \uax obtained from the union graph of Fig.~\ref{f:pathological}. To the left the final partition that cannot be further refined and to the right the corresponding \WA defined according to Definition~\ref{def:finalaut}.}\label{f:uprime}
\end{figure}

Consider again the example of Fig.~\ref{f:pathological}. The partition $\Bx_0 = \myset{s}$, $\Bx_1 = \myset{x^*,x}$, $\Bx_2 = \myset{z}$, $\Bx_3 = \myset{z^*}$, $\Bx_4=\myset{v}$, $\Bx_5=\myset{w^*}$ $\Bx_6=\myset{w}$ $\Bx_7=\myset{y^*,y}$ (right side of Fig.~\ref{f:pathological} and left side of Fig.~\ref{f:uprime}) cannot be further refined. Applying our procedure we obtain the automaton at the right of Fig.~\ref{f:uprime} which is a \WA with the ordering $\Bx_0<\Bx_1<\ldots<\Bx_7$.

Note that the union automaton shown in the left of Fig.~\ref{f:uprime} {\em is not} a \WA: condition~\eqref{eq:wgB} applied to edges $(y^*,z^*)$ and $(y,z)$ together with $z<z^*$ implies $y<y^*$ which in turns implies $x<x^*$ which clashes with $w^* < w$ and the edges $(x^*,w^*)$ and $(x,w)$. This is therefore an instance of a problem where \ua does not have a \WO but our algorithm still returns a \WA. 

\begin{figure}[h]
\centering
  \begin{center}
\includegraphics[width=0.75\columnwidth]{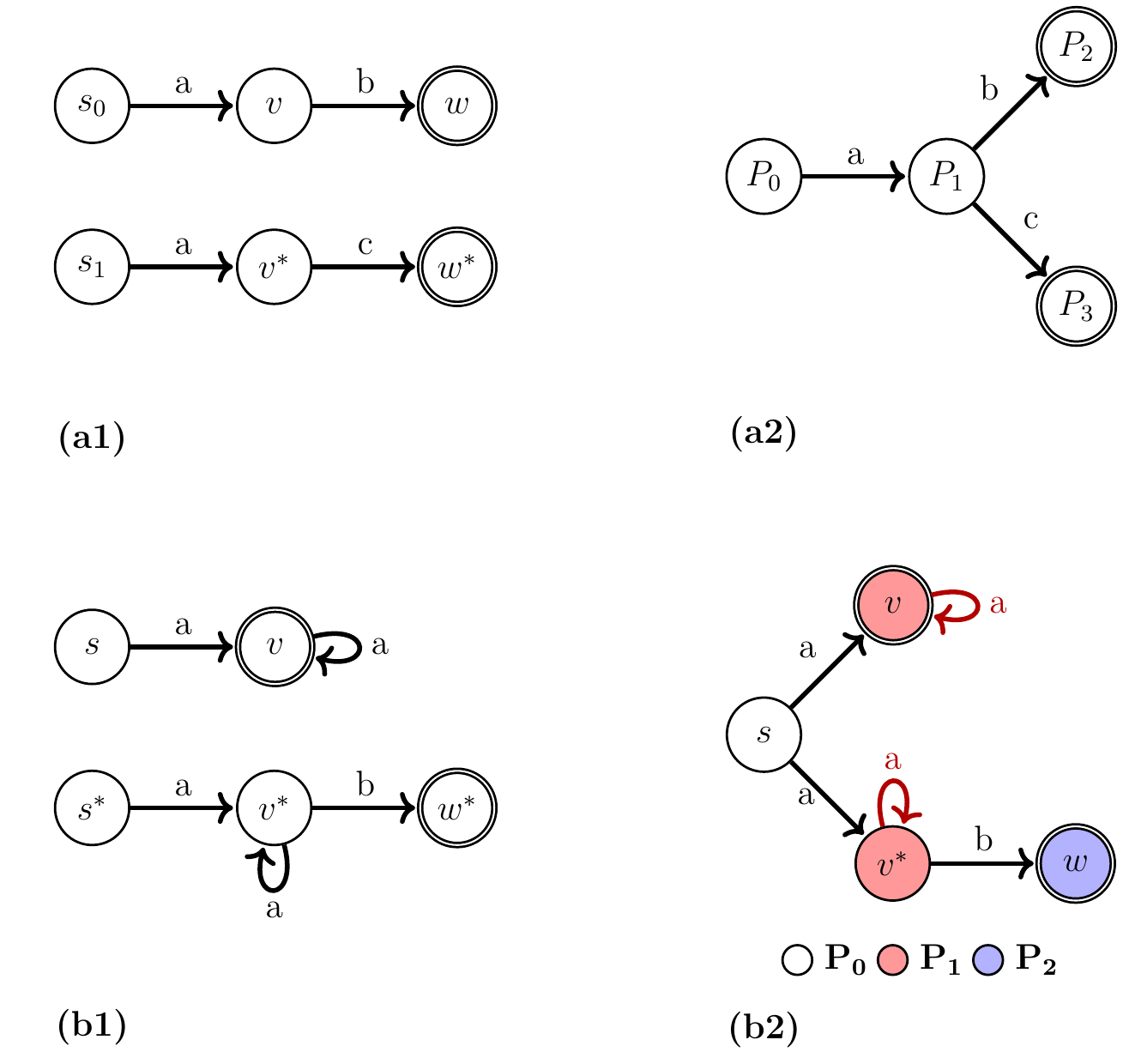}
 \end{center}
  \caption{{\bf Top:} Given the automata on the left {\bf (a1)} the algorithm of Fig.~\ref{fig:refine} returns the automaton on the right {\bf (a2)}, having collapsed the nodes $v$ and $v^*$ to a single node $P_1$. However, the union automaton itself has a \WO, for example $s<v <v^*<w<w^*$. {\bf Bottom:} The union of the two Wheeler automata on the left {\bf (b1)}  is the automaton on the right {\bf (b2)}. After computing the partition $P_0$, $P_1$, $P_2$, our algorithm correctly reports that the union automaton does not have a \WO and stops. However, a \WA for the union language could be obtained for example making $v^*$ a final state in the automaton on the lower left.}
  \label{fig:negex}
\end{figure}

\added{Unfortunately, our algorithm can return a reduced automaton $\uax\neq \ua$ even when $\ua$ does have a \WO, as shown in the example of Fig.~\ref{fig:negex} (top). Although a smaller automaton is preferable for the construction of succinct data structures, this example implies that our algorithm does not provide a complete solution to Problem~\ref{prob:2} since, in case $\uax\neq\ua$, we cannot say whether a \WO for $\ua$ exists.}

\added{With regard to Problem~\ref{prob:1}, 
although our algorithm is correct, in that it returns a Wheeler automaton for the union language, it is not complete since it fails to do so in some cases in which a Wheeler automaton for the union language exists.
This is shown in the example of Fig.~\ref{fig:negex} (bottom) which uses the same automata as in Fig.~\ref{f:automata}. After computing the partition $P_0$, $P_1$, $P_2$ for the nodes of the union automaton, our algorithm reports that it has no \WO, since the \minmax pairs for $v$ and $v^*$ in partition $\Bx_1$ are both $(0,1)$ and thus, by Definition~\ref{def:compat}, are not compatible. However, as noticed in Fig.~\ref{f:automata}, a Wheeler automaton for the union language exists and can be obtained by the automaton on the lower left and making $v^*$ a final state.
Noticing that in Fig.~\ref{fig:negex} {\bf (b2)} $v$ and $v^*$ belong to the same partition $P_1$, this example suggests that a possible strategy for improving our algorithm could be to analyze the union automaton and eliminate redundant nodes before starting the refinement steps.}

\subsection{Implementation details and analysis}\label{sec:merge:alysis}

Let $n_0 = |V_0|$, $m_0=|E_0|$, $n_1 = |V_1|$, $m_1=|E_1|$.  We assume the iterative refining algorithm described in this section takes as input the compact representation of the graphs $G_0$ and $G_1$ supporting constant time navigation operations (see Section~\ref{s:wnotation}). 

The main body of the algorithm is the refinement phase in which the initial \Wcons partition given by Lemma~\ref{lemma:startWcons} is progressively refined.
Let $\Bx_0, \Bx_1, \ldots, \Bx_k$ denote the current partition. We know that $\Bx_0 = \{s\}$ and does not change during the algorithm; for simplicity in our implementation we represent $\Bx_0$ as $\{ s_0, s_1\}$. 
We represent $\Bx_0, \Bx_1, \ldots, \Bx_k$ with two binary arrays $\Bp$ and $\Zp$ both of length $n_0+n_1$. The array $\Bp[1,n_0+n_1]$ encodes the size of the sets: it has exactly $k+1$ \oneb s in positions: $|\Bx_0|$, $|\Bx_0|+|\Bx_1|$, \ldots, $|\Bx_0|+|\Bx_1| + \cdots + |\Bx_k|$. With this encoding for example the size of $\Bx_i$ is given by $\sel_1(\Bp,i+1) - \sel_1(\Bp,i)$, where we assume $\sel_1(\Bp,0)= 0$. The array $\Zp$ encodes the content of each set $\Bx_i$: we logically partition it into $k+1$ subarrays, the $i$-th subarray has length $|\Bx_i|$ and consists of $|\Bx_i \cap V_0|$ \zerob's followed by  $|\Bx_i \cap V_1|$ \oneb's. For example, for the initial partition mentioned in the caption of Fig.~\ref{f:pathological}, recalling that we represent $P_0$ with $\{s_0,s_1\}$ it is
\begin{subequations}
\begin{align*}
\Bp & = \zerob\oneb\, \zerob\zerob\zerob\oneb\, \zerob\zerob\oneb\, \zerob\oneb \\
\Zp & = \zerob\oneb\, \zerob\zerob\oneb\oneb\, \zerob\oneb\oneb\, \zerob\oneb
\end{align*}
\end{subequations}
(spaces have been added for readability). 
The above arrays highlight the similarities between the refining algorithm and the merging algorithm for \dbG graphs. The array $\Zp$ corresponds to the arrays $\bv{h}$ used in Section~\ref{ss:phase1} to denote status of the merging of the \dbG graphs nodes, and  the array $\Bp$ is analogous to the integer array, also called $B$, used in
{the algorithm in Fig.~\ref{fig:xHMalgo}}
to mark block boundaries. Both algorithms compute the merging of the input nodes by iteratively partitioning the very same nodes they are merging.

Assuming we enrich $\Bp$ and $\Zp$ with auxiliary data structures to support constant time $\rank$ and $\sel$ operations, a single refinement operation (Lines~\ref{line:loop0}--\ref{line:loop1} in Fig.~\ref{fig:refine}) is implemented as follows. Setting $b_i = \sel_1(\Bp,i-1)$, we compute the starting position in $\Zp$ of the subarray corresponding to $\Bx_i$ and its length $|\Bx_i| = \sel_1(\Bp,i) - b_i $. Then we compute $|\Bz| = \rank_0(\Zp,b_i + |\Bx_i|) - \rank_0(\Zp,b_i)$ and $|\Bo| = |\Bx_i| - |\Bz|$ (recall that $\Bz = \Bx_i\cap V_0$ and $\Bo = \Bx_i \cap V_1$).

For the merging operation at Line~\ref{line:merge} in Fig.~\ref{fig:refine} we need to compute the \minmax pair for each node in $\Bz$ and $\Bo$. Consider for example the $j$-th node in $\Bz$ (for $\Bo$ it is analogous). Setting $c_j = j+\rank_0(\Zp,b_i)$ we compute the rank of such node in $G_0$; using $G_0$'s succinct representation we compute the largest and smallest nodes, say $\alpha_j$ and $\beta_j$, such that the edges $(\alpha_j,c_j)$ and $(\beta_j,c_j)$ belong to $E_0$ (the computation of $\alpha_j$ is the one outlined just before Lemma~\ref{lemma:WGspace}). The \minmax pair $(\ell,m)$ coincides with the ids of the blocks containing $\alpha_j$ and $\beta_j$, which are given respectively by
$$
\ell = \rank_1(\Bp, \sel_0(\Zp,\alpha_j)),\qquad
m = \rank_1(\Bp, \sel_0(\Zp,\beta_j)).
$$
Note that the \minmax pairs are computed on the spot for the elements that are compared by the merging algorithm so they require only constant storage. The output of the merging is stored into another binary array $\Zpx[1,n_0+n_1]$. The output relative to $\Bx_i$ is written to $\Zpx$ from position $b_i+1$ up to position $b_i + |\Bx_i|$: if the $k$-th element computed by the merging algorithm comes from $\Bz$ we set $\Zpx[b_j+k]=\zerox$ otherwise we set $\Zpx[b_j+k]= \onex$.

Finally, the splitting of $\Bx_i$ (Line~\ref{line:split}) and the update of the current partition (Line~\ref{line:loop1}) is done using another array $\Bpx[1,n_0+n_1]$. For $k=1,\ldots,|Bx_i|-1$ if the \minmax pair corresponding to $\Zpx[b_j+k]$ is different from the one corresponding to $\Zpx[b_j+k+1]$ we set $\Bpx[b_j+k] = \oneb$, otherwise (the \minmax pairs are equal) we set $\Bpx[b_j+k]=\zerob$. We conclude the processing of $\Bx_i$ setting $\Bpx[b_j+|\Bx_i|]=\oneb$. It is immediate to see that at the end of the algorithm of Fig.~\ref{fig:refine} the binary arrays $\Bpx$ and $\Zpx$ provide the encoding of the new partition. If they are equal to $\Bp$ and $\Zp$ then no further refinements are possible and we proceed with the construction of $\uax$, otherwise we replace $\Bp$ and $\Zp$ with $\Bpx$ and $\Zpx$, build the auxiliary data structures supporting rank and select, and proceed with the next iteration. Summing up, the algorithm in Fig.~\ref{fig:refine} takes $\Oh(n_0+n_1)$ time and uses $4(n_0+n_1) + o(n_0+n_1)$ bits of working space. Since the number of refinement steps is at most $\delta+2 = \Oh(|V|)$ the total time is $\Oh(|V|^2)$ and the working space is $4|V| + o(|V|)$ bits.

Given the arrays $\Bp$ and $\Zp$ from the last refinement steps, the succinct representation of the automata $\uax$ (i.e. the arrays $L$, $I$, and $O$)  can be easily computed in $\Oh(|V|+|E|)$ time without using additional working space. We can therefore summarize our result as follows. 

\begin{theorem}
Given the succinct representation of the \WAs $\Auz$ and $\Auo$, our algorithm either reports that the union automaton \ua has no \WO or returns a \WA \uax recognizing the same language as $\ua$. Our algorithm takes $\Oh(|V|^2)$ time and uses $4|V| + o(|V|)$ bits of working space.\qed
\end{theorem}

\section*{Acknowledgments}

\added{
The authors thank some anonymous reviewers for their comments that improved the quality of manuscript, and for pointing out some inconsistencies in some passages of the original draft.}

\paragraph{Funding.}
L.E. and G.M. were partially supported by PRIN grant 2017WR7SHH and by the INdAM-GNCS Project 2020 {\sl MFAIS-IoT}. L.E. was partially supported by the University of Eastern Piedmont project  {\sl HySecEn}.
F.A.L. was supported by the grants $\#$2017/09105-0 and $\#$2018/21509-2 from the S\~ao Paulo Research Foundation (FAPESP).

%
%
\bibliographystyle{splncs04}
\bibliography{bwt}


\end{document}